\documentclass[a4paper,fleqn,usenatbib]{mnras}

\usepackage{newtxtext,newtxmath}

\usepackage[T1]{fontenc}
\usepackage{ae,aecompl}

\usepackage{graphicx}	
\usepackage{amsmath}	
\usepackage{amssymb}	
\usepackage{array}      
\usepackage{booktabs}   
\usepackage{pdflscape}	
\usepackage{multicol}   
\usepackage{xfrac}      

\title[Spiral arm formation mechanisms]{Spiral Structure in Barred galaxies. Observational constraints to spiral arm formation mechanisms}

\author[J. Font et al.]{
Joan Font,$^{1,2}$\thanks{E-mail: jfont@iac.es}
John E. Beckman,$^{1,2,3}$
Phil A. James$^{4}$
and Panos A. Patsis$^{5}$
\\
$^{1}$Instituto de Astrof\'isica de Canarias, 38200, La Laguna, Tenerife , Spain\\
$^{2}$Departamento de Astrof\'isica, Universidad de La Laguna, Tenerife, Spain\\
$^{3}$Consejo Superior de Investigaciones Cient\'ificas, Spain\\
$^{4}$Astrophysics Research Institute, Liverpool John Moores University, IC2, Liverpool Science Park, 146 Brownlow Hill, Liverpool L3 5RF, UK\\
$^{5}$Research Center for Astronomy, Academy of Athens, Soranou Efessiou 4, GR-115 27 Athens, Greece
}

\date{Accepted XXX. Received YYY; in original form ZZZ}

\pubyear{2018}

\begin{document}
\label{firstpage}
\pagerange{\pageref{firstpage}--\pageref{lastpage}}
\maketitle

\begin{abstract}
A method which we have developed for determining corotation radii, has allowed us to map in detail the radial resonant structures of barred spiral galaxies. Here we have combined this information with new determinations of the bar strength and the pitch angle of the innermost segment of the spiral arms to find relationships between these parameters of relevance to the dynamical evolution of the galaxies.  We show how (1) the bar mass fraction, (2) the scaled bar angular momentum, (3) the pitch angle, and (4) the shear parameter vary along the Hubble sequence, and we also plot along the Hubble sequence (5) the scaled bar length, (6) the ratio of bar corotation radius to bar length, (7) the scaled bar pattern speed, and (8) the bar strength. It is of interest to note that the parameters (2), (5), (6), (7), and (8) all show breaks in their behaviour at type Scd. We find that bars with high shear have only small pitch angles, while bars with large pitch angles must have low shear; we also find a generally inverse trend of pitch angle with bar strength.  An inference which at first seems counter-intuitive is that the most massive bars rotate most slowly but have the largest angular momenta. Among a further set of detailed results we pick out here the 2:1 ratio between the number of spiral arms and the number of corotations ouside that of the bar. These results give a guideline to theories of disc-bar evolution.
\end{abstract}




\section{Introduction}

In a series of observational articles \citep{Font2011,Font2014a,Font2014b,Font2017,Beckman2018} we have shown that by using the predicted phase change of the radial gas flow from inwards to outwards, or vice-versa associated with a corotation in a density wave system, (the FB method), it is possible to measure corotation radius with considerable precision and reliability. In the most recent of these articles we confirmed this by comparing the results with those of the Tremaine-Weinberg (TW) method \citep{Tremaine1984} for measuring pattern speeds, and finding complete agreement. The results support very strongly the presence of density waves in galaxy discs. However, we also find corotations associated not only with bars (typically just beyond the end of the bar, both for major bars and for nuclear bars) but also associated with the spiral arms outside the bars, in essentially all the observed galaxies. In general we find more than one corotation radius as we proceed outwards through the part of the disc containing the arms. These kinematic results are clearly important for understanding disc structure, and in the light of the work on arm pitch angles we find it interesting to see how the morphology and the kinematics may be related. To do this we need to look not only at the arms but also at the other structural elements, notably the bar and the bulge. This article is designed to look for quantitative relations between the arm structure, as represented by the pitch angle, the bar mass and angular momentum, the disc mass and angular momentum and, where possible, the bulge mass. 
We selected our sample for this based on our previous kinematic studies of over 100 galaxies in \citet{Font2014a}. In section 2 we describe as a summary of previous studies how we derived the kinematic properties of the arms and the bars in the sample, which are used as observational input data for the present work. We also explain the sources of the images used in the complementary morphological analysis. In section 3 we explain how we measured the morphological and kinematic parameters of the galaxies from the observational data. In section 4 we present the variation of the morphological and kinematic parameters, including the pitch angle, with Hubble morphological class, and in section 5 we conclude by suggesting how the rotational properties of the galaxies can be related, via their morphology, to their evolutionary histories. 

\subsection{Pitch Angles}
The aim of this article is to combine information about the pitch angles of spiral arms in disc galaxies with kinematic information related to their resonant structure, to improve our understanding of the processes involved in arm formation and evolution. Historically spiral arms have been described by logarithmic spirals, although, as pointed out by \citet{Kennicutt1981} hyperbolic spirals may also be used, because for many galaxies their arms are not so long (\(\Delta\theta < 360^\circ\)) as to make the difference between the two functions significant. The pitch angle is defined as the angle between the tangent to a spiral arm at a given point and the tangent to a circle centred on the galaxy centre at the same point. According to this definition, the pitch is 0$^{\circ}$ when the circle and the spiral are coincident, and can reach an extreme value of 90$^{\circ}$ when they are perpendicular. Small values of the pitch angle thus correspond to tightly wound arms, and large values to more open arms. With this definition of the pitch angle, a logarithmic spiral is conveniently expressed by 

\begin{equation}
    r(\theta) = r_0 e^{\theta \tan(\varphi)}
	\label{eq:spiral}
\end{equation}

where \(r\) and $\theta$ are standard polar coordinates, \(r_0\) is the radius at the starting point taken as \(\theta=0^{\circ}\) and $\varphi$ is the pitch angle. A variety of methods to calculate the pitch angle has been described in the literature. One of the most used, introduced by \citet{Kalnajs1975}, and later developed by \citet{Considere1982,Considere1988}, \citet{Puerari1992, Puerari2000}, \citet{Seigar2008} and \citet{Mata-Chavez2014} consists of performing a two-dimensional Fourier transform decomposition of the intensity distribution on the basis of logarithmic spirals. Average values for pitch angles over complete galaxies were derived this way in \citet{Rio1999} for a sample of nine spirals, in \citet{Davis2012} for a sample of 49 galaxies, while in \citet{Savchenko2013} the method was used on 50 galaxies to obtain an average value of the pitch angle as a function of galactocentric radius. Another Fourier based technique, using a 1D transform of azimuthal profiles, was applied by \citet{Grosbol2004} to determine several parameters of the spiral arms of 54 galaxies: their relative amplitudes, radial extent, and pitch angles. An alternative method uses the slope of the arm more directly \citep{Seigar1998} to derive $\varphi$ using a version of Equation~\ref{eq:spiral} in which the pitch angle is expressed directly:

\begin{equation}
    \tan(\varphi) = \left| \frac{d(\ln{r})}{d\theta} \right|
	\label{eq:pitch}
\end{equation}

\citet{Ringermacher2009} proposed a more general version of this method, which can be applied for different shapes, such as ring galaxies with arms which can even spiral inwards as well as outwards.  Modern variants of the above methods include that by \citet{Puerari2014} based on the correlations between circular windows in the ($\theta$, ln\,$r$) plane and logarithmic spiral arms with different pitch angles. This method was used to analyze the pitch angle as a function of scale and position for a couple of grand design galaxies. Another approach was developed by \citet{Shields2015} who obtained the pitch angle for a sample of 30 galaxies, by determining the best fit to a galaxy image with a set of spiral templates of known pitch angle. \citet{Hayes2012} and \citet{Davis2014} developed a code using computer vision techniques to identify independent segments of the spiral arms and used the algorithm to extract the value of the pitch angle and the length of each segment, among other parameters. This was then applied to a large sample, of 644,000 galaxies from the Sloan survey. They found that though a constant value for the pitch angle was a satisfactory assumption for each segment, it changes significantly from segment to segment. This result shows generally that the pitch angle is not a constant as a function of galactocentric radius, which is relevant for the work presented here.

The pitch angles of the spiral arms have been used as parameters to explore   physical structures. In a brief review \citet{Kennefick2014} outlined different ways in which measuring the pitch angle can be related to other properties, and can therefore, under suitable conditions, be used as a proxy for them. These include the bulge mass, \citep{Seigar2008} and, given the relation between bulge mass and central black hole mass, the pitch angle can also be related to the latter (see \citet{Davis2017}. In a more general way the pitch angle can be used in tests of theories which show how the spiral arms are themselves formed. In an early article by \citet{Schlosser1984} they compared the predictions of spiral arm formation in the  density wave and stochastic self-propagation theories using their own  pitch angle determinations and those of \citet{Kennicutt1981} concluding that agreement with the second scenario is considerably better. The general prediction that it should be possible to test for long-term stability in spiral arms using the systematic change in colour across the arm was tested observationally by \citet{Gonzalez1996, Grosbol2006, Tamburro2008, Egusa2009, Martinez2009, Foyle2011, Cedres2013}; \citet{Foyle2011} concluded that this was not observed, so that stable arms seemed to be precluded. The relevance of the pitch angle in this context was referred to in \citet{Kennefick2014} who pointed out that linear density wave theory implies a tight relationship between pitch angle, disc density, and bulge mass, whereas swing amplification for spiral arm production does not.

\section{The observational data}

In \citet{Font2014a} we measured the resonant structure, i.e. the resonance radii and pattern speeds of a sample of 104 galaxies, applying the Font-Beckman method \citep{Font2011, Beckman2018} (FB). Later, in \citet{Font2017} we performed a set of morphological and kinematical measurements, which include the bar length, the bar strength, the corotation radius of the bar and the bar pattern speed, of a subsample of 68 barred galaxies and studied the variation of these parameters with the morphological type of the galaxy, and also how each parameter is related to the others. In the present study we extend this latter sample to 79 barred spiral galaxies in the local Universe (z < 0.03), and for each we combine different geometric measurements with kinematic measurements of the bar and the spiral structure.

We performed precise measurements of the corotation radii of all the galaxies and their corresponding pattern speeds using FB, which requires a 2D radial velocity field. For the present study we have used a high spatial and spectral resolution Fabry-P\'erot data-cube in the H$\alpha$ emission line, which consists of a two-dimensional image with spatial coordinates (\textit{x,y}) and a third spectral dimension, which can be calibrated in wavelength or velocity, and so for each pixel of the data-cube we obtain an H$\alpha$ line profile from which we can determine the position of the peak and hence build a velocity map. The majority of the galaxies of our sample, 74 of a total set of 79, were observed with the GHASP instrument \citep[Galaxy HAlpha survey of SPirals,][]{Epinat2008} at the 1.93\,m telescope of the Observatoire de Haute Provence in France, during the period 1998-2004. The GHASP Fabry-Perot produced data cubes with a pixel scale of 0.68\,arcsec/pixel, a spectral sampling resolution of $\sim$\,16\,km\,s$^{-1}$ in H$\alpha$, and an angular resolution limited by the seeing with an average value of $\sim$\,3\,arcsec. All spectroscopic data, including the data cubes and the moment maps, are available online\footnote{\url{https://cesam.lam.fr/fabryperot/}}.
The five remaining galaxies (UGC3013, UGC5303, UGC5981, UGC6118 and UGC7420) were observed with GH$\alpha$FaS \citep[Galaxy H$\alpha$ Fabry-P\'erot System,][]{Hernandez2008}. The observations were carried out during several runs at the William Herschel Telescope, Roque de los Muchachos Observatory, La Palma, Spain, in the period between 2010 and 2014, with an average seeing value of $\sim$\,1.0\,arcsec. This Fabry-P\'erot interferometer produces data cubes of 3.4\,arcmin$^2$ in angular size with a pixel scale of 0.192\,arcsec/pixel and a sampling spectral resolution of $\sim$\,6\,km\,s$^{-1}$ in H$\alpha$.

In order to perform the measurements of the bar length and the bar strength, and the measurements of the pitch angle of the spiral arms, as well as to perform the decomposition of the galaxy to determine the relative mass contributions of the bulge, the bar and the disc, we used images from different surveys depending on their availability and quality. We gave priority to infrared images, so most images of the galaxies of our sample are 3.6\,$\mu$m infrared images taken from the Spitzer archive\footnote{\url{http://sha.ipac.caltech.edu/applications/Spitzer/SHA/}}. For the particular case of UGC2855, which is not found in the Spitzer survey, we used the infrared image in the J band from the 2MASS survey\footnote{\url{https://irsa.ipac.caltech.edu/applications/2MASS/PubGalPS/}}, as the quality is good enough to perform the corresponding calculations. A small subset of seven galaxies is not available in any infrared survey, and so for four of them (UGC9465, UGC9736, UGC9969 and UGC11557) we used images from the Sloan Digital Sky Survey in the r band (data release 12 of the SDSS III\footnote{\url{http://www.sdss.org/dr12/}}), while for the remaining three galaxies (UGC2080, UGC11124, and UGC12276) the images are taken from the ESO Digitized Sky
Survey\footnote{\url{http://archive.eso.org/dss/dss}} (DSS2-red), as this is the only survey that provides data with high enough resolution to make reliable calculations for these objects.

\begin{table*}
    \centering
    \caption{Properties of Galaxies. Column (1) identifies the galaxy using the UGC classification; the galaxies are also named according the conventional NGC and IC classifications in Column (2). Columns (3) and (4) give the morphological type according to RC3 and \citet{Buta2015}, respectively. In column (5) appears the distance of the object, and column (6) gives the radius for the 25 B-band mag\,arcsec$^{-2}$ isophote according to the NED database. Columns (7) and (8) show, respectively, the values of the inclination angle and the position angle of the line of nodes of the galaxy. The asymptotic rotational velocity determined from the rotation curves is listed in column (9).}
    \label{tab:table1}
    \begin{tabular}{*9c} 
	    \hline
        \multicolumn{2}{c}{Name} &  \multicolumn{2}{c}{Morphology} & D & r$_{25}$ & \textit{i} & P.A. & v$_{asym}$\\
        UGC & NGC & \multicolumn{2}{c}{} & (Mpc) & (arcsec) & ($^{\circ}$) & ($^{\circ}$) & (km\,s$^{-1}$)\\
        (1) & (2) & (3) & (4) & (5) & (6) & (7) & (8) & (9)\\
        \hline
	    089 & 023 & SB(s)a & - & 51.44 & 62.7  & 33 & 177 & 330\\
	    508 & 266 & SB(rs)ab & - & 63.8 & 88.55 & 25 & 123 & 530\\
	    763 & 428 & SAB(s)m & SAB(s)dm & 12.7 & 122.2 & 54 & 117 & 104\\
        1256 & 672 & SB(s)cd & (R?)SB(s)d & 7.2 & 217.35 & 76 & 73 & 85\\
        1317 & 674 & SAB(r)c & - & 42.2 & 134 & 73 & 106 & 205\\
        1437 & 753 & SAB(rs)bc & - & 66.8 & 75.35 & 47 & 307 & 218\\
        1736 & 864 & SAB(rs)c & SAB(s)bc & 17.6 & 140.3 & 35 & 27 & 193\\
        1913 & 925 & SAB(s)d & - & 9.3 & 314.15 & 48 & 288 & 105\\
        2080 & IC 239 & SAB(rs)cd  & - & 13.7 & 137.15 & 25 & 336 & 131\\
        2855 & - & SABc & - & 17.5 & 130.95 & 68 & 100 & 229\\
        3013 & 1530 & SB(rs)b & - & 37.0 & 137.15 & 55 & 195 & 212\\
        3463 & IC 2166 & SAB(s)bc & - & 38.6 & 90.6 & 63 & 110 & 168\\
        3685 & - & SB(rs)b & - & 26.3 & 99.35 & 12 & 298 & 102\\
        3709 & 2342 & S pec & - & 70.7 & 41.4 & 55 & 232 & 230\\
        3740 & 2276 & SAB(rs)c & - & 17.1 & 84.55 & 48 & 247 & 87\\
        3809 & 2336 & SAB(r)bc & - & 32.9 & 212.4 & 58 & 357 & 258\\
        3915 & IC 2199 & SBbc & - & 66.0 & 32.9 & 47 & 30 & 200\\
        4165 & 2500 & SB(rs)d  & SAB(s)d & 11.0 & 86.5 & 41 & 265 & 80\\
        4273 & 2543 & SB(s)b & SAB(s)b & 35.4 & 70.35 & 60 & 212 & 200\\
        4325 & 2552 & SA(s)m & (R')SAB(s)m & 10.9 & 104 & 63 & 57 & 85\\
        4422 & 2595 & SAB(rs)c & - & 58.1 & 94.85 & 25 & 36 & 345\\
        4555 & 2649 & SAB(rs)bc & - & 58.0 & 47.55 & 38 & 90 & 185\\
        4936 & 2805 & SAB(rs)d & (R)SA(s)c pec & 25.6 & 189.3 & 13 & 294 & 230\\
        5228 & - & SB(s)c & (R$_2$')SAB(s)bc & 24.7 & 73.65 & 72 & 120 & 125\\
        5303 & 3041 & SAB(rs)c & SA(rs)c & 17.7 & 111.45 & 36 & 273 & 202\\
        5316 & 3027 & SB(rs)d & SB(s)dm & 16.14 & 127.95 & 71 & 130 & 95\\
        5319 & 3061 & (R')SB(rs)c & SAB(rs)b pec & 35.8 & 49.8 & 30 & 345 & 180\\
        5351 & 3067 & SAB(s)ab & SB(s)dm / Sph & 21.32 & 73.65 & 65 & 219 & 130\\
        5510 & 3162 & SAB(rs)bc & SA(s)bc & 18.6 & 90.6 & 31 & 200 & 167\\
        5532 & 3147 & SA(rs)bc & SAB(rs)b & 41.1 & 116.7 & 32 & 147 & 398\\
        5786 & 3310 & SAB(r)bc pec & SA(rs)bc pec & 14.2 & 92.7 & 53 & 153 & 80\\
        5840 & 3344 & (R)SAB(r)bc & SAB(r)bc & 6.9 & 212.4 & 25 & 333 & 251\\
        5842 & 3346 & SB(rs)cd & SB(rs)cd & 15.2 & 86.5 & 47 & 292 & 110\\
        5981 & 3433 & SA(s)c & 	SAB(rs)b & 32.44 & 106.45 & 38.5 & 294 & 206\\
        5982 & 3430 & SAB(rs)c & SAB(r)bc & 20.8 & 106.5 & 55 & 28 & 199\\
        6118 & 3504 & (R)SAB(s)ab & (R$_1$')SAB(r,nl)a & 19.8 & 80.75 & 19 & 330 & 240\\
        6277 & 3596 & SAB(rs)c & SA(s)bc & 17.55 & 119.45 & 17 & 76 & 275\\
        6537 & 3726 & SAB(r)c & SAB(r)bc & 14.3 & 185 & 47 & 200 & 187\\
        6778 & 3893 & SAB(rs)c & SA(s)c & 15.5 & 134 & 49 & 343 & 223\\
        7021 & 4045 & SAB(r)a & (R$_1$'L)SAB(r,nl)ab & 26.8 & 80.75 & 56 & 266 & 175\\
        7154 & 4145 & SAB(rs)d & SAB(rs)d & 16.2 & 176.65 & 65 & 275 & 145\\
        7323 & 4242 & SAB(s)dm & (L)IAB(s)m & 8.1 & 150.35 & 51 & 38 & 84\\
        7420 & 4303 & SAB(rs)bc & SAB(rs,nl)c & 20.0 & 193.7 & 29 & 135 & 177\\
        7766 & 4559 & SAB(rs)cd & SB(s)cd & 13.0 & 321.45 & 69 & 323 & 120\\
        7853 & 4618 & SB(rs)m & (R')SB(rs)m & 8.9 & 125.05 & 58 & 217 & 62\\
        7876 & 4635 & SAB(s)d & SA(s)d & 14.5 & 61.25 & 53 & 344 & 98\\
        7985 & 4713 & SAB(rs)d & SAB(rs)cd & 13.7 & 80.75 & 49 & 276 & 112\\
        8403 & 5112 & SB(rs)cd & SB(s)cd & 19.1 & 119.45 & 57 & 121 & 120\\
        8709 & 5297 & SAB(s)c & SAB$_x$(s)bc sp & 35.0 & 168.7 & 76 & 330 & 207\\
        8852 & 5376 & SAB(r)b & - & 30.6 & 62.7 & 52 & 63 & 186\\
        8937 & 5430 & SB(s)b & (R$_1$')SB(s,nl)b & 49.0 & 65.65 & 32 & 185 & 275\\
        9179 & 5585 & SAB(s)d & - & 5.7 & 172.65 & 36 & 49 & 111\\
        9358 & 5678 & SAB(rs)b & (R$_1$'L)SAB(rs)b pec & 29.1 & 99 & 54 & 182 & 221\\
        9366 & 5668 & SA(rs)bc & SAB(rs)c & 37.7 & 119.45 & 62 & 225 & 241\\
        9465 & 5727 & SABdm & - & 26.4 & 67.15 & 65 & 127 & 97\\
        9576 & 5774 & SAB(rs)d & - & 23.23 & 90.6 & 41 & 122 & 108\\
        9649 & 5832 & SB(rs)b &	SAB(s)m & 9.29 & 111.45 & 54 & 235 & 98\\

	    \hline
    \end{tabular}
\end{table*}

\begin{table*}
    \centering
    \contcaption{Properties of Galaxies}
    \label{tab:cont_table1}
    \begin{tabular}{*9c} 
	    \hline
        \multicolumn{2}{c}{Name} &  \multicolumn{2}{c}{Morphology} & D & r$_{25}$ & \textit{i} & P.A. & v$_{asym}$\\
        UGC & NGC & \multicolumn{2}{c}{} & (Mpc) & (arcsec) & ($^{\circ}$) & ($^{\circ}$) & (\,km\,s$^{-1}$)\\
        (1) & (2) & (3) & (4) & (5) & (6) & (7) & (8) & (9)\\
        \hline
        9736 & 5874 & SAB(rs)c & - & 45.4 & 68.75 & 51 & 219 & 192\\
        9753 & 5879 & SA(rs)bc & SAB(rs)bc & 12.4 & 125.05 & 69 & 3 & 138\\
        9943 & 5970 & SB(r)c  & SB(s)c & 28.0 & 86.5 & 54 & 266 & 185\\
        9969 & 5985 & SAB(r)b & SAB(s)ab & 36.0 & 164.85 & 61 & 16 & 311\\
        10075 & 6015 & SA(s)cd & SAB$_a$(s)cd & 14.7 & 161.1 & 62 & 210 & 168\\
        10359 & 6140 & SB(s)cd pec & SB(s)d & 16.0 & 189.3 & 44 & 284 & 143\\
        10470 & 6217 & (R)SB(rs)bc & (R')SB(rs)b & 21.2 & 90.6 & 34 & 287 & 164\\
        10546 & 6236 & SAB(s)cd & SB(s)dm & 20.4 & 86.5 & 42 & 182 & 106\\
        10564 & 6248 & SBd & - & 18.4 & 94.85 & 77 & 149 & 75\\
        11012 & 6503 & SA(s)cd & SAB(s)bc & 5.3 & 212.4 & 72 & 299 & 117\\
        11124 & - & SB(s)cd & - & 23.7 & 75.35 & 51 & 182 & 96\\
        11283 & IC 1291 & SB(s)dm & - & 31.3 & 54.6 & 34 & 120 & 173\\
        11407 & 6764 & SB(s)bc & - & 35.8 & 68.75 & 64 & 65 & 158\\
        11429 & 6792 & SBb  & - & 66.75 & 67.15 & 61 & 208 & 200\\
        11498 & - & SBb & - & 43.54 & 94.85 & 71 & 251 & 282\\
        11557 & - & SAB(s)dm & - & 19.7 & 65.65 & 29 & 276 & 105\\
        11597 & 6946 & SAB(rs)cd & - & 5.54 & 344.45 & 40 & 241 & 155\\
        11861 & - & SABdm & - & 25.1 & 104 & 43 & 218 & 181\\
        11872 & 7177 & SAB(r)b & - & 18.1 & 92.7 & 47 & 86 & 183\\
        12276 & 7440 & SB(r)a & - & 77.8 & 42.4 & 33 & 322 & 94\\
        12343 & 7479 & SB(s)c & - & 26.9 & 122.2 & 52 & 203 & 221\\
        12754 & 7741 & SB(s)cd & (R$_2$')SB(s)cd & 8.9 & 130.95 & 53 & 342 & 123\\

	    \hline
    \end{tabular}
\end{table*}

\section{Data Analysis}
In this section we give a detailed description of how we measured the morphological and kinematical parameters of the galaxies. The values of all the parameters measured in this study also include values for their uncertainties. We classify all parameters reported in the present article into three types: Type A parameters are those that are taken from other data bases, in which case uncertainties are not always available. Parameters of type B are those which we measure from an image or a velocity map by applying a specific method; the uncertainties for these parameters are not straightforward to determine and depend on the method used, so we give an estimate of such uncertainties taking into account the resolution of the image (angular resolution) or map (angular and spectral resolutions) used, the uncertainty of the parameters needed in the method, and the uncertainties introduced by the method itself. Finally the parameters of type C are calculated from any other type of parameter following a given expression, which is used to compute the associated uncertainties by applying the uncertainty propagation technique.

\subsection{The basic parameters of each galaxy}
We give two essential parameters when describing each galaxy, the morphological type and the distance of the galaxy in Mpc. The latter is taken from the NASA Extragalactic Database (NED); the values are given in column 5 of the Table~\ref{tab:table1}, and are used to calculate the conversion factor between angular and spatial measurements. The morphology of each galaxy is also obtained from the same database and is given according to the Bright Galaxy Catalogue, RC3 \citep{Vaucouleurs1995} in column 3 of Table~\ref{tab:table1}. To complement the morphological information of the galaxies, we additionally provide, in column 4 of Table~\ref{tab:table1}, the morphological type according to the classifications of \citet{Buta2015}, who made a detailed classification of 2352 nearby galaxies in the S$^4$G survey \citep[Spitzer Survey of Stellar Structure in Galaxies,][]{Sheth2010}, using 3.6\,$\mu$m and 4.5\,$\mu$m infrared images.

\subsection{The parameters of the disc}
In order to characterize each galactic disc, we determined several parameters of the three types, according to the parameter classification described above.

Parameters of type A. Within this category, we obtained the values of r$_{25}$, defined as the radius corresponding to the 25 B-band mag\,arcsec$^{-2}$ isophote, from the RC3 catalogue \citep{Vaucouleurs1995}, which is available in the NASA Extragalactic Database. In this case, uncertainties are not provided in the RC3 catalogue, so we assume an average uncertainty for all galaxies of $\pm$\,5.4\,arcsec, which is the average uncertainty for this parameter according to the Hyperleda database\footnote{\url{http://leda.univ-lyon1.fr/}}. The values for r$_{25}$ are given in Table~\ref{tab:table1}, column 6.

Parameters of type B. We calculated several parameters of this category, such as the position of the centre of the galaxy, the inclination angle of the disc galaxy, the position angle of the line of nodes, and the asymptotic circular velocity. The asymptotic velocity is defined as the maximum rotation velocity of the ionised gas, which in practical terms is the velocity to which the rotation curve tends at large galactocentric radii, where this curve is almost flat. The rotation curve is derived using the ROTCUR task of the GIPSY astronomical software package\footnote{\url{https://www.astro.rug.nl/~gipsy/}}, which performs fits of tilted rings to the velocity map at different radii. In doing this, all the geometrical parameters, which include the position of the galactic centre, the inclination angle, and the position angle of the major axis are also calculated by allowing one of these parameters to vary freely while the others are fixed, thus yielding the variation of this parameter with galactocentric radius.  We take an average value of that parameter in those rings which contain more pixels and show less dispersion of the fitted parameter; this is repeated with the next parameter, and so on. The values of the inclination angle, the position angle and the asymptotic rotation velocity are shown in columns 7, 8 and 9 of Table~\ref{tab:table1}. Concerning the uncertainties of these parameters, we have estimated that all values of the inclination angle have an uncertainty of 7\,$^{\circ}$, and the uncertainty in the position angle is 2\,$^{\circ}$, while for the maximum circular velocity, a fractional uncertainty of 10\,\% is assumed.

\begin{figure}
	\includegraphics[width=\columnwidth]{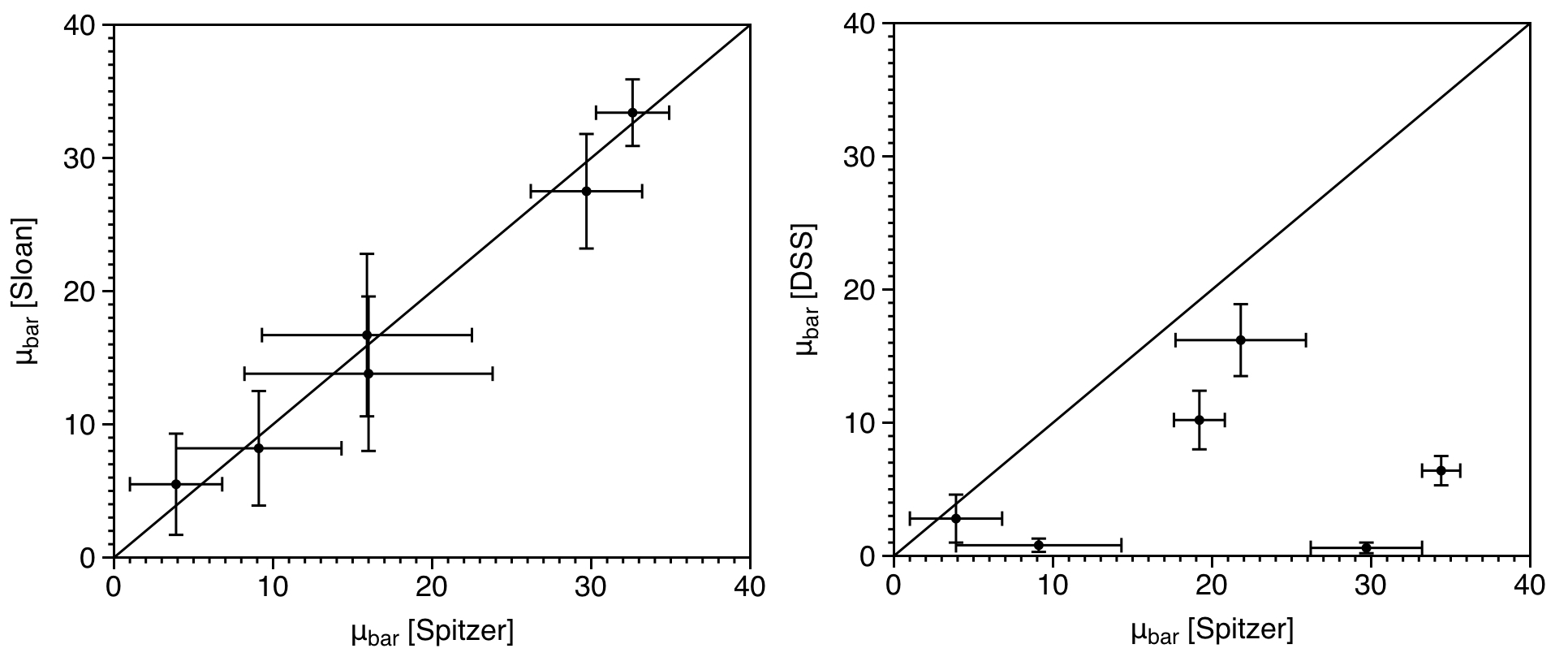}
    \caption{Comparison of the mass fraction of the bar calculated from images of different surveys. (Left panel) Results using near-infrared images are compared with those from \textit{Sloan} in the r band. (Right panel) The same as the left panel, but using DSS images. The dashed line in the two panels plots the 1:1 relation.}
    \label{fig:mu_comparative}
\end{figure}

Another B-type parameter we calculated for the disc galaxy is the stellar mass fraction of the disc. To do this, we have used the publicly available DiskFit software\footnote{\url{https://www.physics.queensu.ca/Astro/people/Kristine\_Spekkens/diskfit
/}} \citep{Sellwood2015}. Given an image of the galaxy (preferably a near-infrared image) and initial reliable values of several morphological parameters describing three structures of the galaxy: the disc (position angle, ellipticity), the bar (position angle, ellipticity) and the bulge (effective radius, Sersic index, position angle and ellipticity), the code performs fittings of multi-component models and by minimizing a $\chi^2$ estimate between the model and the image, it provides the contribution of the bar, the bulge and the disc to the galaxy light, including the associated uncertainties, along with the best fitting values of the morphological parameters. We estimated the goodness of the image decomposition by comparing the fitting values of the position angle and the inclination angle of the disc, and the position angle of the bar, with those provided in the GHASP database or calculated by ellipse fitting. While the code produces reliable fits when the Sloan images in the R band are used, the fitting parameters calculated are unrealistic with a DSS image; this is illustrated in Fig.~\ref{fig:mu_comparative}, where we plot, for a small subset of six objects, the bar mass fraction derived from the Spitzer image against the bar mass fraction calculated from the Sloan R band image (left panel), and against this parameter computed with the DSS image (right panel), in the two figures the solid black line marks the 1:1 relation. In consequence, we do not give the disc mass fraction for the three DSS galaxies in our sample (i.e. UGC2080, UGC11124 and UGC12276). The mass fraction of the disc and its uncertainty appear in Table~\ref{tab:table2}, column 2.

Parameters of type C. As the only parameter of this type, we computed the angular velocity of the outer disc, $\omega_{disc}$, which is defined as the asymptotic rotation velocity, v$_{asym}$ (in km\,s$^{-1}$), divided by the isophotal radius at the 25 B-band mag\,arcsec$^{-2}$, r$_{25}$ (in kpc). The values found for this parameter together with their uncertainties are given in column 3 of Table~\ref{tab:table2}. The angular velocity of the outer disc is used to calculate a set of scaled parameters, which are described in the following subsections.

\begin{table*}
    \centering
    \caption{Parameters of the disc, the bar and the spiral arms. Column (1) identifies the galaxy using the UGC classification; Columns (2) and (3) give the stellar mass fraction and the angular velocity of the disc, respectively. From column (4) to column (8), parameters characterising the bar are given: the bar length, the bar strength, the mass fraction, the bar pattern speed, and the relative angular momentum of the bar. Two parameters describe the spiral arms: the pattern speed in column (9), and the pitch angle in column (10). Uncertainties associated with the pattern speeds, given in columns (7) and (9), are the mean values of the upper and lower uncertainties.}
    \label{tab:table2}
    \begin{tabular}{c c c c c c c c c c} 
	    \hline
        Name &  $\mu_{disc}$ & $\omega_{disc}$ & r$_{bar}$ & S$_{b}$ & $\mu_{bar}$ & $\Omega_{bar}$ & $\lambda_{bar}$ & $\Omega_{spiral}$ & $\varphi$ \\
        UGC & (\%) & (\,km\,s$^{-1}$\,kpc$^{-1}$) & (arcsec) & & (\%) & (\,km\,s$^{-1}$\,kpc$^{-1}$) & ($\times10^{-3}$) & (\,km\,s$^{-1}$\,kpc$^{-1}$) &  ($^{\circ}$)\\
        (1) & (2) & (3) & (4) & (5) & (6) & (7) & (8) & (9) & (10)\\
        \hline
        089 & 79.4 $\pm$\,9.4 & 21.1 $\pm$\,9.5 & 6.2 $\pm$\,0.6 & 0.62 $\pm$\,0.05 & 17.6 $\pm$\,8.3 & 163.2 $\pm$\,4.3 & 5.6 $\pm$\,2.7 & 97.1 $\pm$\,0.4 & 13.7 $\pm$\,7.2\\
508 & 63.9 $\pm$\,4.7 & 19.4 $\pm$\,1.5 & 48.2 $\pm$\,1.4 & 0.37 $\pm$\,0.03 & 21.8 $\pm$\,4.1 & 32.6 $\pm$\,1.2 & 56.9 $\pm$\,13.9 & 26.5 $\pm$\,2.3 & 10.7 $\pm$\,5.5\\
763 & 69.4 $\pm$\,7.5 & 13.8 $\pm$\,0.2 & 43.5 $\pm$\,3.2 & 0.30 $\pm$\,0.03 & 28.2 $\pm$\,6.3 & 26.9 $\pm$\,0.6 & 32.5 $\pm$\,9.5 & 13.8 $\pm$\,0.7 & 19.4 $\pm$\,4.5\\
1256 & 88.7 $\pm$\,11.1 & 11.2 $\pm$\,1.0 & 41.3 $\pm$\,16.1 & 0.17 $\pm$\,0.03 & 11.3 $\pm$\,3.6 & 26.5 $\pm$\,1.0 & 3.6 $\pm$\,3.0 & 21.9 $\pm$\,0.6 & 30.6 $\pm$\,15.5\\
1317 & 94.4 $\pm$\,2.7 & 7.5 $\pm$\,0.5 & 16.4 $\pm$\,0.8 & 0.45 $\pm$\,0.03 & 2.0 $\pm$\,1.5 & 34.7 $\pm$\,5.4 & 0.5 $\pm$\,0.3 & 15.5 $\pm$\,0.2 & 19.5 $\pm$\,5.5\\
1437 & 93.2 $\pm$\,20.3 & 8.9 $\pm$\,0.4 & 11.1 $\pm$\,0.1 & 0.10 $\pm$\,0.03 & 3.4 $\pm$\,1.6 & 51.7 $\pm$\,5.6 & 1.5 $\pm$\,0.8 & 9.6 $\pm$\,0.3 & 11.6 $\pm$\,4.0\\
1736 & 88.8 $\pm$\,1.3 & 16.1 $\pm$\,0.7 & 43.1 $\pm$\,12.8 & 0.15 $\pm$\,0.03 & 7.4 $\pm$\,1.9 & 33.5 $\pm$\,0.9 & 5.4 $\pm$\,3.6 & 22.5 $\pm$\,2.8 & 21.9 $\pm$\,2.0\\
1913 & 40.6 $\pm$\,8.2 & 7.4 $\pm$\,0.8 & 106.3 $\pm$\,10.9 & 0.22 $\pm$\,0.03 & 31.72 $\pm$\,14.1 & 19.8 $\pm$\,0.8 & 79.7 $\pm$\,42.7 & 19.8 $\pm$\,0.6 & 18.9 $\pm$\,3.5\\
2080 & - & 14.4 $\pm$\,1.3 & 26.6 $\pm$\,3.1 & 0.35 $\pm$\,0.03 & - & 46.1 $\pm$\,1.3 & - & 46.1 $\pm$\,2.3 & 13.6 $\pm$\,2.5\\
2855 & 82.7 $\pm$\,2.6  & 20.6 $\pm$\,1.8 & 43.4 $\pm$\,0.6 & 0.30 $\pm$\,0.03 & 7.2$\pm$\,0.6 & 33.5 $\pm$\,2.9 & 5.2 $\pm$\,0.6 & 24.2 $\pm$\,0.4 & 25.5 $\pm$\,3.5\\
3013 & 22.4 $\pm$\,6.9 & 8.6 $\pm$\,0.3 & 88.7 $\pm$\,16.3 & 1.21 $\pm$\,0.03 & 33.6 $\pm$\,4.9 & 11.4 $\pm$\,0.3 & 276.7 $\pm$\,139.8 & 10.4 $\pm$\,0.4 & 11.1 $\pm$\,6.5\\
3463 & 66.7 $\pm$\,6.3 & 9.9 $\pm$\,0.9 & 30.2 $\pm$\,1.8 & 0.25 $\pm$\,0.03 & 20.3 $\pm$\,3.5 & 19.1 $\pm$\,2.8 & 21.8 $\pm$\,5.9 & 13.3 $\pm$\,0.3 & 18.3 $\pm$\,4.5\\
3685 & 78.0 $\pm$\,1.7 & 8.1 $\pm$\,0.2 & 26.4 $\pm$\,3.2 & 0.40 $\pm$\,0.03 & 13.3 $\pm$\,0.4 & 22.3 $\pm$\,0.3 & 11.1 $\pm$\,2.8 & 18.8 $\pm$\,0.7 & 14.6 $\pm$\,3.0\\
3709 & 72.7 $\pm$\,1.2 & 16.2 $\pm$\,1.5 & 19.2 $\pm$\,2.5 & 0.54 $\pm$\,0.03 & 11.2 $\pm$\,3.5 & 27.0 $\pm$\,1.6 & 18.4 $\pm$\,7.6 & 27.0 $\pm$\,0.5 & 34.9 $\pm$\,3.0\\
3740 & 80.7 $\pm$\,1.2 & 12.4 $\pm$\,0.1 & 18.3 $\pm$\,1.1 & 0.30 $\pm$\,0.03 & 5.6 $\pm$\,1.2 & 21.5 $\pm$\,0.5 & 1.9 $\pm$\,0.5 & 14.4 $\pm$\,1.4 & 31.5 $\pm$\,1.5\\
3809 & 55.3 $\pm$\,2.9 & 7.6 $\pm$\,0.4 & 47.4 $\pm$\,5.7 & 0.12 $\pm$\,0.03 & 25.9 $\pm$\,2.1 & 24.6 $\pm$\,1.7 & 25.2 $\pm$\,6.8 & 12.3 $\pm$\,0.1 & 13.2 $\pm$\,2.5\\
3915 & 80.9 $\pm$\,2.0 & 19.0 $\pm$\,2.2 & 14.8 $\pm$\,1.5 & 0.23 $\pm$\,0.05 & 9.76 $\pm$\,1.6 & 42.0 $\pm$\,5.8 & 35.6 $\pm$\,9.5 & 23.9 $\pm$\,0.5 & 29.5 $\pm$\,8.0\\
4165 & 73.7 $\pm$\,9.0 & 17.3 $\pm$\,0.3 & 36.8 $\pm$\,4.3 & 0.34 $\pm$\,0.03 & 9.11 $\pm$\,5.2 & 27.3 $\pm$\,1.0 & 11.8 $\pm$\,7.5 & 18.5 $\pm$\,1.9 & 31.2 $\pm$\,3.0\\
4273 & 62.4 $\pm$\,1.6 & 16.6 $\pm$\,0.3 & 38.4 $\pm$\,6.2 & 0.25 $\pm$\,0.03 & 20.89 $\pm$\,0.3 & 22.4 $\pm$\,1.2 & 44.9 $\pm$\,14.8 & 14.7 $\pm$\,0.5 & 26.7 $\pm$\,4.5\\
4325 & 86.4 $\pm$\,11.4 & 15.5 $\pm$\,0.6 & 60.4 $\pm$\,1.7 & 0.22 $\pm$\,0.03 & 13.13 $\pm$\,12.0 & 24.1 $\pm$\,0.6 & 12.6 $\pm$\,11.7 & 18.5 $\pm$\,1.2 & 41.6 $\pm$\,2.0\\
4422 & 55.9 $\pm$\,2.7 & 12.9 $\pm$\,1.0 & 41.1 $\pm$\,2.5 & 0.55 $\pm$\,0.03 & 31.61 $\pm$\,1.7 & 35.1 $\pm$\,1.0 & 42.5 $\pm$\,8.3 & 29.2 $\pm$\,1.8 & 26.8 $\pm$\,3.0\\
4555 & 94.9 $\pm$\,12.3 & 13.8 $\pm$\,1.9 & 9.9 $\pm$\,0.6 & 0.15 $\pm$\,0.03 & 0.49 $\pm$\,0.2 & 41.1 $\pm$\,5.6 & 0.22 $\pm$\,0.10 & 22.3 $\pm$\,1.1 & 14.2 $\pm$\,2.5\\
4936 & 63.0 $\pm$\,2.7 & 9.8 $\pm$\,0.2 & 40.6 $\pm$\,2.4 & 0.22 $\pm$\,0.03 & 18.0 $\pm$\,5.7 & 24.1 $\pm$\,0.6 & 10.8 $\pm$\,3.7 & 16.3 $\pm$\,0.5 & 17.3 $\pm$\,3.0\\
5228 & 69.8 $\pm$\,8.6 & 14.2 $\pm$\,0.7 & 24.4 $\pm$\,6.7 & 0.45 $\pm$\,0.03 & 28.63 $\pm$\,8.4 & 32.3 $\pm$\,1.1 & 34.3 $\pm$\,21.8 & 21.0 $\pm$\,0.5 & 24.1 $\pm$\,5.0\\
5303 & 70.2 $\pm$\,9.3 & 21.1 $\pm$\,1.0 & 26.2 $\pm$\,1.6 & 0.12 $\pm$\,0.03 & 9.4 $\pm$\,4.9 & 56.9 $\pm$\,8.9 & 6.7 $\pm$\,3.8 & 29.6 $\pm$\,0.4 & 26.2 $\pm$\,3.0\\
5316 & 88.2 $\pm$\,8.1 & 9.5 $\pm$\,0.4 & 45.8 $\pm$\,6.2 & 0.72 $\pm$\,0.03 & 11.42 $\pm$\,10.8 & 12.0 $\pm$\,0.4 & 7.0 $\pm$\,3.5 & 11.0 $\pm$\,0.5 & 14.9 $\pm$\,8.9\\
5319 & 90.3 $\pm$\,0.8 & 20.8 $\pm$\,2.2 & 11.8 $\pm$\,1.4 & 0.30 $\pm$\,0.03 & 6.60 $\pm$\,1.1 & 51.3 $\pm$\,2.4 & 3.3 $\pm$\,1.0 & 29.8 $\pm$\,2.7 & 22.7 $\pm$\,3.5\\
5351 & 84.8 $\pm$\,6.9 & 17.1 $\pm$\,5.1 & 23.1 $\pm$\,2.7 & 0.42 $\pm$\,0.03 & 10.60 $\pm$\,3.3 & 57.5 $\pm$\,4.9 & 13.8 $\pm$\,5.6 & 45.3 $\pm$\,0.6 & 16.78 $\pm$\,6.9\\
5510 & 73.4 $\pm$\,1.5 & 20.4 $\pm$\,0.8 & 16.1 $\pm$\,1.4 & 0.49 $\pm$\,0.03 & 11.84 $\pm$\,1.8 & 50.9 $\pm$\,2.0 & 4.2 $\pm$\,1.1 & 32.4 $\pm$\,2.7 & 30.5 $\pm$\,2.5\\
5532 & 68.3 $\pm$\,4.4 & 17.1 $\pm$\,8.1 & 9.0 $\pm$\,2.1 & 0.13 $\pm$\,0.03 & 1.60 $\pm$\,0.2 & 115.9 $\pm$\,22.6 & 0.3 $\pm$\,0.1 & 54.1 $\pm$\,0.5 & 11.8 $\pm$\,2.5\\
5786 & 41.5 $\pm$\,3.0 & 12.5 $\pm$\,0.9 & 21.2 $\pm$\,1.9 & 0.49 $\pm$\,0.03 & 28.67 $\pm$\,3.0 & 22.2 $\pm$\,7.7 & 21.9 $\pm$\,6.9 & 13.4 $\pm$\,1.2 & 15.4 $\pm$\,3.5\\
5840 & 54.7 $\pm$\,11.7 & 35.3 $\pm$\,0.9 & 32.5 $\pm$\,2.7 & 0.12 $\pm$\,0.03 & 15.93 $\pm$\,6.6 & 84.6 $\pm$\,6.2 & 5.4 $\pm$\,2.7 & 57.9 $\pm$\,1.0 & 17.7 $\pm$\,4.0\\
5842 & 91.2 $\pm$\,1.0 & 17.3 $\pm$\,1.6 & 18.7 $\pm$\,3.3 & 0.17 $\pm$\,0.03 & 5.34 $\pm$\,0.6 & 39.0 $\pm$\,0.8 & 2.1 $\pm$\,0.7 & 21.8 $\pm$\,1.4 & 18.3 $\pm$\,5.5\\
5981 & 84.3 $\pm$\,2.6 & 12.3 $\pm$\,2.2 & 15.4 $\pm$\,1.2 & 0.15 $\pm$\,0.05 & 9.18 $\pm$\,2.2 & 46.5 $\pm$\,3.6 & 2.9 $\pm$\,0.8 & 29.5 $\pm$\,0.3 & 11.8 $\pm$\,5.5\\
5982 & 84.8 $\pm$\,11.7 & 16.5 $\pm$\,0.9 & 15.2 $\pm$\,1.2 & 0.17 $\pm$\,0.03 & 11.15 $\pm$\,3.4 & 73.3 $\pm$\,4.5 & 3.1 $\pm$\,1.2 & 20.7 $\pm$\,0.5 & 17.0 $\pm$\,4.0\\
6118 & 37.3 $\pm$\,1.9 & 31.0 $\pm$\,1.9 & 38.6 $\pm$\,5.3 & 0.57 $\pm$\,0.03 & 19.14 $\pm$\,2.3 & 58.9 $\pm$\,1.9 & 287.8 $\pm$\,90.1 & 58.9 $\pm$\,1.5 & 18.8 $\pm$\,5.5\\
6277 & 84.9 $\pm$\,1.9 & 27.1 $\pm$\,2.1 & 11.6 $\pm$\,0.5 & 0.13 $\pm$\,0.05 & 9.02 $\pm$\,1.6 & 115.3 $\pm$\,12.3 & 1.4 $\pm$\,0.3 & 61.4 $\pm$\,0.6 & 22.7 $\pm$\,11.0\\
6537 & 59.4 $\pm$\,7.0 & 14.6 $\pm$\,0.5 & 49.1 $\pm$\,4.7 & 0.24 $\pm$\,0.03 & 35.66 $\pm$\,5.7 & 35.4 $\pm$\,0.8 & 34.3 $\pm$\,9.6 & 20.9 $\pm$\,0.7 & 16.1 $\pm$\,5.2\\
6778 & 77.6 $\pm$\,9.3 & 22.1 $\pm$\,1.2 & 19.2 $\pm$\,1.4 & 0.12 $\pm$\,0.03 & 13.83 $\pm$\,2.7 & 65.0 $\pm$\,5.7 & 3.4 $\pm$\,0.9 & 31.5 $\pm$\,0.7 & 17.4 $\pm$\,3.5\\
7021 & 54.8 $\pm$\,5.6 & 16.7 $\pm$\,2.1 & 27.3 $\pm$\,2.0 & 0.86 $\pm$\,0.03 & 13.43 $\pm$\,5.5 & 48.2 $\pm$\,6.2 & 26.9 $\pm$\,12.5 & 34.4 $\pm$\,0.8 & 23.1 $\pm$\,3.8\\
7154 & 79.4 $\pm$\,1.8 & 10.5 $\pm$\,0.2 & 41.6 $\pm$\,3.1 & 0.71 $\pm$\,0.03 & 19.22 $\pm$\,1.6 & 20.0 $\pm$\,0.5 & 8.5 $\pm$\,1.5 & 9.5 $\pm$\,0.3 & 18.5 $\pm$\,3.9\\
7323 & 73.6 $\pm$\,11.2 & 14.2 $\pm$\,0.2 & 59.0 $\pm$\,2.6 & 0.47 $\pm$\,0.03 & 25.9 $\pm$\,11.8 & 18.6 $\pm$\,0.9 & 23.6 $\pm$\,11.8 & 15.0 $\pm$\,1.3 & 18.5 $\pm$\,3.8\\
7420 & 41.7 $\pm$\,15.6 & 9.4 $\pm$\,0.4 & 33.8 $\pm$\,2.3 & 0.44 $\pm$\,0.03 & 31.11 $\pm$\,15.0 & 49.6 $\pm$\,3.0 & 15.0 $\pm$\,8.6 & 30.4 $\pm$\,0.2 & 18.3 $\pm$\,3.2\\
7766 & 82.3 $\pm$\,2.8 & 5.9 $\pm$\,0.2 & 22.9 $\pm$\,5.0 & 0.17 $\pm$\,0.03 & 15.31 $\pm$\,3.5 & 39.4 $\pm$\,1.5 & 2.1 $\pm$\,1.0 & 19.4 $\pm$\,0.2 & 12.3 $\pm$\,3.0\\
7853 & 92.5 $\pm$\,15.3 & 11.5 $\pm$\,0.5 & 33.4 $\pm$\,10.3 & 0.27 $\pm$\,0.03 & 6.6 $\pm$\,3.1 & 19.0 $\pm$\,0.5 & 2.8 $\pm$\,2.1 & 16.6 $\pm$\,3.2 & 25.3 $\pm$\,4.3\\
7876 & 80.7 $\pm$\,2.8 & 22.8 $\pm$\,0.9 & 18.8 $\pm$\,2.3 & 0.20 $\pm$\,0.03 & 2.96 $\pm$\,1.4 & 36.4 $\pm$\,1.3 & 1.8 $\pm$\,0.9 & 28.6 $\pm$\,1.6 & 35.3 $\pm$\,1.8\\
7985 & 92.8 $\pm$\,5.7 & 20.9 $\pm$\,2.4 & 20.5 $\pm$\,0.9 & 0.29 $\pm$\,0.03 & 4.68 $\pm$\,2.1 & 38.3 $\pm$\,4.0 & 2.0 $\pm$\,0.9 & 22.4 $\pm$\,1.2 & 17.1 $\pm$\,4.2\\
8403 & 75.6 $\pm$\,2.5 & 10.8 $\pm$\,0.4 & 18.0 $\pm$\,7.4 & 0.32 $\pm$\,0.03 & 23.21 $\pm$\,2.2 & 20.2 $\pm$\,0.4 & 4.3 $\pm$\,2.5 & 14.4 $\pm$\,0.5 & 26.4 $\pm$\,2.1\\
8709 & 61.0 $\pm$\,3.1 & 7.2 $\pm$\,0.8 & 38.7 $\pm$\,5.7 & 0.54 $\pm$\,0.03 & 23.77 $\pm$\,2.5 & 24.1 $\pm$\,0.5 & 22.8 $\pm$\,7.2 & 12.8 $\pm$\,0.2 & 25.6 $\pm$\,2.0\\
8852 & 86.8 $\pm$\,5.5 & 20.0 $\pm$\,1.4 & 19.7 $\pm$\,2.3 & 0.11 $\pm$\,0.03 & 2.83 $\pm$\,5.8 & 44.5 $\pm$\,1.8 & 2.4 $\pm$\,1.6 & 37.4 $\pm$\,0.5 & 17.4 $\pm$\,2.6\\
8937 & 43.7 $\pm$\,5.9 & 17.6 $\pm$\,1.6 & 28.0 $\pm$\,7.7 & 1.06 $\pm$\,0.03 & 36.10 $\pm$\,3.0 & 35.9 $\pm$\,3.7 & 101.7 $\pm$\,62.3 & 23.8 $\pm$\,3.4 & 17.7 $\pm$\,2.1\\
9179 & 77.7 $\pm$\,13.0 & 23.3 $\pm$\,0.2 & 69.9 $\pm$\,8.2 & 0.30 $\pm$\,0.03 & 3.90 $\pm$\,2.9 & 55.8 $\pm$\,1.3 & 1.0 $\pm$\,0.8 & 30.3 $\pm$\,3.7 & 13.8 $\pm$\,4.3\\
9358 & 53.1 $\pm$\,8.5 & 15.8 $\pm$\,3.0 & 12.9 $\pm$\,4.4 & 0.37 $\pm$\,0.03 & 9.31 $\pm$\,2.6 & 92.4 $\pm$\,9.2 & 5.8 $\pm$\,4.4 & 39.4 $\pm$\,0.5 & 18.0 $\pm$\,2.0\\
9366 & 72.6 $\pm$\,14.5 & 11.03 $\pm$\,1.5 & 26.8 $\pm$\,1.4 & 0.25 $\pm$\,0.03 & 7.95 $\pm$\,2.2 & 34.1 $\pm$\,4.9 & 5.7 $\pm$\,2.1 & 26.9 $\pm$\,0.2 & 15.0 $\pm$\,3.8\\
9465 & 95.0 $\pm$\,12.3 & 11.3 $\pm$\,0.5 & 11.8 $\pm$\,1.4 & 0.20 $\pm$\,0.03 & 4.8 $\pm$\,2.3 & 26.7 $\pm$\,1.2 & 1.2 $\pm$\,0.6 & 14.7 $\pm$\,0.5 & 23.5 $\pm$\,3.5\\
9576 & 97.5 $\pm$\,0.1 & 10.6 $\pm$\,0.8 & 10.5 $\pm$\,2.6 & 0.32 $\pm$\,0.05 & 2.67 $\pm$\,0.1 & 26.5 $\pm$\,1.9 & 0.3 $\pm$\,0.1 & 16.0 $\pm$\,0.4 & 10.2 $\pm$\,4.5\\
9649 & 91.0 $\pm$\,10.8 & 19.5 $\pm$\,0.8 & 44.4 $\pm$\,2.0 & 0.17 $\pm$\,0.05 & 8.66 $\pm$\,6.5 & 46.2 $\pm$\,1.9 & 11.9 $\pm$\,9.1 & 31.3 $\pm$\,0.9 & 10.3 $\pm$\,4.4\\
        
	    \hline
    \end{tabular}
\end{table*}

\begin{table*}
    \centering
    \contcaption{Properties of the disc, the bar and the spiral arms}
    \label{tab:cont_table2}
    \begin{tabular}{c c c c c c c c c c} 
	    \hline
        Name &  $\mu_{disc}$ & $\omega_{disc}$ & r$_{bar}$ & S$_{b}$ & $\mu_{bar}$ & $\Omega_{bar}$ & $\lambda_{bar}$ & $\Omega_{spiral}$ & $\varphi$ \\
        UGC & (\%) & (\,km\,s$^{-1}$\,kpc$^{-1}$) & (arcsec) & & (\%) & (\,km\,s$^{-1}$\,kpc$^{-1}$) & ($\times10^{-3}$) & (\,km\,s$^{-1}$\,kpc$^{-1}$) &  ($^{\circ}$)\\
        (1) & (2) & (3) & (4) & (5) & (6) & (7) & (8) & (9) & (10)\\
        \hline
9736 & 95.8 $\pm$\,0.9 & 12.7 $\pm$\,0.7 & 8.6 $\pm$\,0.4 & 0.34 $\pm$\,0.03 & 2.1 $\pm$\,0.4 & 46.6 $\pm$\,6.1 & 0.4 $\pm$\,0.1 & 14.5 $\pm$\,0.5 & 13.2 $\pm$\,1.8\\
9753 & 50.5 $\pm$\,3.5 & 18.4 $\pm$\,1.2 & 27.3 $\pm$\,11.4 & 0.22 $\pm$\,0.03 & 39.01 $\pm$\,2.6 & 74.6 $\pm$\,4.8 & 49.9 $\pm$\,31.5 & 34.9 $\pm$\,0.6 & 11.0 $\pm$\,2.6\\
9943 & 64.2 $\pm$\,1.0 & 15.8 $\pm$\,0.5 & 19.4 $\pm$\,3.1 & 0.27 $\pm$\,0.03 & 26.09 $\pm$\,2.4 & 39.5 $\pm$\,4.3 & 17.0 $\pm$\,6.1 & 19.5 $\pm$\,0.4 & 26.1 $\pm$\,3.2\\
9969 & 82.7 $\pm$\,7.5 & 10.8 $\pm$\,0.8 & 17.5 $\pm$\,1.6 & 0.42 $\pm$\,0.03 & 16.65 $\pm$\,8.1 & 50.2 $\pm$\,2.5 & 3.5 $\pm$\,1.8 & 16.3 $\pm$\,0.2 & 12.5 $\pm$\,3.5\\
10075 & 87.1 $\pm$\,2.3 & 14.6 $\pm$\,0.2 & 12.3 $\pm$\,2.0 & 0.17 $\pm$\,0.03 & 2.37 $\pm$\,0.5 & 59.5 $\pm$\,3.3 & 0.21 $\pm$\,0.08 & 18.9 $\pm$\,0.4 & 18.4 $\pm$\,3.4\\
10359 & 89.2 $\pm$\,3.0 & 9.7 $\pm$\,0.5 & 15.8 $\pm$\,4.6 & 0.50 $\pm$\,0.03 & 9.79 $\pm$\,1.8 & 36.7 $\pm$\,0.6 & 1.0 $\pm$\,0.5 & 14.8 $\pm$\,1.0 & 23.5 $\pm$\,3.4\\
10470 & 60.5 $\pm$\,2.5 & 17.6 $\pm$\,0.5 & 45.0 $\pm$\,9.8 & 0.92 $\pm$\,0.03 & 15.89 $\pm$\,2.0 & 24.9 $\pm$\,1.4 & 30.5 $\pm$\,14.4 & 22.0 $\pm$\,2.1 & 16.1 $\pm$\,2.5\\
10546 & 92.5 $\pm$\,2.5 & 12.4 $\pm$\,0.6 & 14.6 $\pm$\,0.9 & 0.10 $\pm$\,0.03 & 4.68 $\pm$\,2.0 & 36.3 $\pm$\,1.5 & 1.4 $\pm$\,0.6 & 18.8 $\pm$\,1.3 & 20.4 $\pm$\,2.5\\
10564 & 90.8 $\pm$\,8.9 & 8.9 $\pm$\,0.3 & 37.5 $\pm$\,4.7 & 0.67 $\pm$\,0.03 & 7.74 $\pm$\,6.2 & 12.4 $\pm$\,0.3 & 6.2 $\pm$\,0.5 & 10.6 $\pm$\,0.5 & 12.3 $\pm$\,5.0\\
11012 & 94.8 $\pm$\,1.8 & 21.4 $\pm$\,1.3 & 17.3 $\pm$\,0.4 & 0.35 $\pm$\,0.03 & 0.37 $\pm$\,0.2 & 107.7 $\pm$\,2.2 & 0.04 $\pm$\,0.02 & 71.8 $\pm$\,0.8 & 11.8 $\pm$\,2.8\\
11124 & - & 11.1 $\pm$\,0.3 & 27.0 $\pm$\,4.3 & 0.30 $\pm$\,0.03 & - & 14.4 $\pm$\,0.3 & - & 13.2 $\pm$\,0.9 & 22.6 $\pm$\,1.5\\
11283 & 52.6 $\pm$\,18.8 & 20.9 $\pm$\,2.3 & 24.0 $\pm$\,2.1 & 0.72 $\pm$\,0.03 & 33.41 $\pm$\,14.1 & 30.7 $\pm$\,2.2 & 59.9 $\pm$\,37.2 & 21.9 $\pm$\,4.4 & 29.8 $\pm$\,1.6\\
11407 & 49.4 $\pm$\,5.8 & 13.2 $\pm$\,0.3 & 52.9 $\pm$\,3.3 & 0.40 $\pm$\,0.03 & 23.32 $\pm$\,9.9 & 13.0 $\pm$\,0.3 & 44.9 $\pm$\,13.3 & 12.9 $\pm$\,1.3 & 20.4 $\pm$\,3.8\\
11429 & 71.0 $\pm$\,1.6 & 9.2 $\pm$\,0.4 & 29.9 $\pm$\,7.9 & 0.64 $\pm$\,0.05 & 18.42 $\pm$\,1.5 & 14.3 $\pm$\,0.7 & 12.7 $\pm$\,4.3 & 11.3 $\pm$\,0.3 & 17.6 $\pm$\,7.8\\
11498 & 77.0 $\pm$\,1.7 & 14.1 $\pm$\,0.8 & 10.3 $\pm$\,1.5 & 0.59 $\pm$\,0.05 & 9.27 $\pm$\,1.7 & 50.9 $\pm$\,2.7 & 1.7 $\pm$\,0.5 & 31.2 $\pm$\,0.3 & 10.5 $\pm$\,5.5\\
11557 & 90.3 $\pm$\,3.8 & 16.7 $\pm$\,0.3 & 23.8 $\pm$\,2.1 & 0.22 $\pm$\,0.03 & 8.94 $\pm$\,3.6 & 18.4 $\pm$\,0.3 & 0.5 $\pm$\,0.3 & 16.3 $\pm$\,5.7 & 39.6 $\pm$\,1.6\\
11597 & 43.4 $\pm$\,5.2 & 16.8 $\pm$\,2.4 & 40.9 $\pm$\,3.0 & 0.32 $\pm$\,0.05 & 24.02 $\pm$\,7.6 & 63.4 $\pm$\,4.3 & 1.0 $\pm$\,0.3 & 46.1 $\pm$\,0.5 & 27.6 $\pm$\,8.0\\
11861 & 90.2 $\pm$\,5.3 & 14.3 $\pm$\,0.6 & 30.8 $\pm$\,2.3 & 0.33 $\pm$\,0.03 & 6.76 $\pm$\,1.4 & 24.8 $\pm$\,1.1 & 3.8 $\pm$\,1.0 & 18.2 $\pm$\,1.5 & 28.9 $\pm$\,2.5\\
11872 & 45.1 $\pm$\,6.4 & 22.5 $\pm$\,4.5 & 18.2 $\pm$\,1.0 & 0.77 $\pm$\,0.03 & 19.16 $\pm$\,1.3 & 94.5 $\pm$\,10.7 & 23.1 $\pm$\,5.2 & 65.2 $\pm$\,0.7 & 16.5 $\pm$\,5.2\\
12276 & - & 5.9 $\pm$\,0.4 & 16.2 $\pm$\,1.8 & 0.17 $\pm$\,0.03 & - & 11.9 $\pm$\,0.6 & - & 9.7 $\pm$\,1.1 & 16.4 $\pm$\,2.1\\
12343 & 47.5 $\pm$\,1.8 & 13.9 $\pm$\,1.0 & 85.9 $\pm$\,22.4 & 0.94 $\pm$\,0.03 & 34.41 $\pm$\,1.2 & 18.4 $\pm$\,1.0 & 158.5 $\pm$\,83.6 & 18.4 $\pm$\,0.4 & 23.3 $\pm$\,1.5\\
12754 & 69.2 $\pm$\,1.5 & 21.8 $\pm$\,0.3 & 62.1 $\pm$\,14.9 & 0.50 $\pm$\,0.03 & 28.91 $\pm$\,2.9 & 36.6 $\pm$\,2.3 & 52.7 $\pm$\,26.1 & 28.4 $\pm$\,1.0 & 14.1 $\pm$\,4.3\\
        
	    \hline
    \end{tabular}
\end{table*}

\subsection{The properties of the bar}
\label{sec:bar}
There is no parameter characterizing the bar that was obtained from a database, so all bar parameters in this study are measured here (type B) or calculated (type C). Some of these parameters were already determined in a previous article \citep{Font2017}, and a detailed description of how we measured these parameters, in addition to alternative methods and definitions of these specific parameters, can be found there; therefore, here we give only a brief description of the method we used to determine the properties of the bar.

Properties of type B. We measured four parameters of B-type concerning the bar: the mass fraction of the bar, the bar length, the bar strength, and the pattern speed of the bar. The bar mass fraction, as described in the previous subsection, is calculated using the Diskfit code for all galaxies with the exception of those for which only the DSS image is available (see Fig.~\ref{fig:mu_comparative}). The calculated values are listed in column 6 of Table~\ref{tab:table2}. The bar length values given in column 4 of Table~\ref{tab:table2} are calculated as the average between the deprojected values of $a_{\epsilon}$ and $a_{bar}$, which are obtained following the precepts of \cite{Erwin2003}; the technique consists of performing ellipse fitting to the image of the galaxy, so the dependence of the ellipticity, and the position angle of the fitting ellipse, on the radius are obtained. The former radius, $a_{\epsilon}$, which is a lower limit of the bar length, is defined as the radius where the ellipticity reaches a maximum for uniform values of the position angle \citep{Wozniak1991}, and the latter radius, $a_{bar}$, which is an upper limit, is determined as the lesser value the radius, outside $a_{\epsilon}$, where the position angle of the fitted ellipses changes by 10$^{\circ}$, and the radius, just outside $a_{\epsilon}$, where the ellipticity shows a local minimum. The deprojection of these radii is calculated using the expression,

\begin{equation}
    a^{deproj} = a_{proj} \cdot \cos \theta_{bar}\cdot(\tan^2 \theta_{bar}\cdot\sec^2 \textit{i}+1)^{^1/_2}
	\label{eq:deprojection}
\end{equation}

where $a_{proj}$ is the projected value obtained from the ellipse fitting, $\theta_{bar}$ is the position angle of the bar with respect to the major axis of the disk, and \textit{i} is the inclination angle of the galaxy. equation~(\ref{eq:deprojection}) is also used to calculate the uncertainties in $a_{\epsilon}$ and $a_{bar}$, which are combined with the uncertainty of the mean in order to determine the uncertainty in $r_{bar}$ given in Table~\ref{tab:table2}.

The bar strength is defined as the maximum tangential force divided by the mean axisymmetric radial force, \(^{F_T^{max}}/_{\langle F_R \rangle}\). In the present study, we calculated the bar strength from the Fourier decomposition of the galaxy image according to the expression

\begin{equation}
    S_b=\frac{\displaystyle\sum_{m=2,4,6}^{} \int_{r_1}^{r_2} \sqrt{A_m^2+B_m^2}dr}{\int_{r_1}^{r_2} A_0dr}
	\label{eq:barstrength}
\end{equation}

where \(A_m\) and \(B_m\) are the harmonic coefficients as functions of the radius, and \(A_0\) is the coefficient of the zero order term (\(m=0\)). These coefficients, together with the associated uncertainty  in each, are calculated using the \textit{kinemetry} code \citep{Krajnovic2006}, which performs a Fourier decomposition of the image of the galaxy, presenting the surface brightness image as a combination of a finite number of harmonic terms. Although the largest contribution to the bar strength comes from the amplitude of the term \(m=2\), \citet{Ohta1990} showed that higher order even terms (\(m=4,6\)) should also be taken into account. The arbitrary integration limits, \(r_1\) and \(r_2\) in equation~(\ref{eq:barstrength}) are values of the radius that characterize the bar region, in order to exclude any contribution of the spiral arm torques to the bar strength. In our calculation we took \(r_1\) to be half of $a_{\epsilon}$ and \(r_2\) to be $a_{bar}$, as these parameters are a lower and an upper limit of the length of the bar (as explained in the previous paragraph). The corresponding uncertainties are calculated by propagating the uncertainties associated with each harmonic coefficient according to equation~(\ref{eq:barstrength}). The bar strength along with the uncertainty can be found in Table~\ref{tab:table2}, column 5.

The corotation radius and the pattern speed are essential parameters that help to describe the dynamics of a galaxy. In the present study, we have applied the Font-Beckman method \citep{Font2011,Font2014a}, which gives precise measurements of the corotation radii of the galaxy, using high resolution velocity fields of the ionized gas, which are produced with a Fabry-Perot interferometer, as we showed in \citet{Beckman2018}, where we compared this method with the Tremaine-Weinberg method for the galaxy NGC3433, obtaining values of the corotation radius (or pattern speed) in good agreement not only for the stellar component but also for the ionized gas component. Density wave theory \citep{Lin1964} predicts that the non-tangential velocity experiences a flip in sign at the radius where corotation occurs \citep{Kalnajs1978}; based on this property, the method identifies those pixels (or bins, when the angular resolution is taken into account) in the residual velocity map, which have a phase change of $\pi$ in the radial component of stellar or gas motion. From the coordinates of the phase-reversals, their galactocentric radii are calculated, deriving the radial distribution of the phase-reversals, which is well organized in separated peaks \citep[see histograms in][]{Font2011}. In general, the Lagrangian points L1,L2, L4, and L5, which define the corotation, are not aligned in a single circle \citep[see figure 3.14 of][]{Binney2008}, this means that we should use the term corotation region rather than corotation radius, and this explains why we obtain peaks in the histogram of the phase-reversals; each peak in the histogram is associated with a different corotation, and by determining the centre and the FWHM of the peak, we obtain the corotation radius and its uncertainty, so we use these two quantities to describe the corotation region. The fact that we find more than one corotation in any galaxy indicates the co-existence of different density waves, each of them characterized by its corresponding pattern speed. Among the measured peaks in the histogram of phase-reversals, we associate the bar corotation with the strongest peak that is centred in a radial position just outside the bar end. In order to quantify how much faster are bars compared to the galactic disk, we define a parameter of type C, $\Gamma_{bar}$, which is the pattern speed of the density wave associated with the bar divided by the angular rate of the disk. The values of the pattern speed and its associated uncertainties can be found in column 7 of Table~\ref{tab:table2}.

Knowing the values of the bar relative mass, the bar length and the bar pattern speed, we calculated two extra parameters of type C: the relative moment of inertia of the bar, $\iota_{bar}$, and the relative bar angular momentum, $\lambda_{bar}$. Assuming that the galactic bar is rotating as a solid bar with respect to its centre, the moment of inertia of the bar with respect to an axis perpendicular to the plane containing the bar can be calculated as

\begin{equation}
    I_{bar}=\frac{1}{3}m_{bar}(r^2_{bar}+w^2_{bar})
	\label{eq:inertia}
\end{equation}

with $m_{bar}$ being the mass of the bar, $r_{bar}$ and $w_{bar}$ the radial extent and the half-width of the bar, respectively. These two latter parameters are related by means of the ellipticity, $\epsilon$, which is defined as \(\epsilon=1-w_{bar}/r_{bar}\). In order to calculate this parameter of the bar relative to the galaxy disc, the bar mass is scaled by the disc mass, and the bar length is scaled by $r_{25}$. Thus, the relative moment of inertia of the bar is calculated using the expression

\begin{equation}
    \iota_{bar} \simeq \frac{m_{bar}}{m_{disc}} f \Big(\frac{r_{bar}}{r_{25}}\Big)^2 = \frac{\mu_{bar}}{\mu_{disc}} f \rho^2_{bar}
	\label{eq:inertia2}
\end{equation}

where the geometrical factor $f$ is defined as \(f=(1+(1-\epsilon)^2)\), and $\mu_{bar}$ and $\mu_{disc}$ are the contribution to the total stellar mass of the bar and the disc, respectively. The angular momentum of the bar is defined as the moment of inertia multiplied by the angular speed of the bar, where the former is calculated following equation~\ref{eq:inertia}, and the latter is the pattern speed of the bar, $\Omega_{bar}$. Rather than using absolute values of these parameters, but instead using disk-scaled values, we calculated the relative bar angular momentum as

\begin{equation}
    \lambda_{bar}=\iota_{bar} \frac{\Omega_{bar}}{\omega_{disk}} = \iota_{bar} \Gamma_{bar}
	\label{eq:angular}
\end{equation}

where $\iota_{bar}$ is the relative bar moment of inertia calculated according equation~\ref{eq:inertia2}, and $\Gamma_{bar}$ is the relative pattern speed of the bar (listed in column 7 of Table~\ref{tab:table2}). The values of the relative angular momentum of the bar obtained using equation~\ref{eq:angular} are given in Table~\ref{tab:table2}, column 8. The associated uncertainties are determined by propagation using the corresponding expressions.

\subsection{The properties of the spiral arms}

In this study we measured two parameters that characterize the spiral arms; the pattern speed associated with the spiral arms, and the spiral pitch angle. These two parameters are of type B as they are obtained applying the Font-Beckman method and the slope method, respectively. 

\citet{Kennicutt1981} and \citet{Savchenko2013} showed that the pitch angle varies with galactocentric radius, so the morphology of the arms should be taken into account in order to measure this parameter. This raises two different options: A. To measure the variation of the pitch angle as a function of the radius, giving, for instance, a set of values of this parameter for a given radial or azimuthal sector of the spiral arm. B. To calculate a single averaged value of the pitch angle. \citet{Davis2014} showed the need to fragment the spiral arms into segments in which the pitch angle remains constant. Based on this idea, we measured the pitch angle of our galaxies performing linear fits in the \((\theta,\ln{r})\) plane for well-defined radial region around the bar corotation radius. Within this region \citep[or segment, adopting the terminology of][]{Davis2014}, the pitch angle is approximately constant, so the slope method is applicable giving reliable measurements of the pitch angle. The value of the pitch angle measured following this technique does not characterize the whole of the spiral arms of a given galaxy, as we do not take into account breaks in the pitch angle nor the branching of the spiral arms.

Despite problems, such as a bias due to the presence of background objects, foreground stars, star forming regions in the arms etc. that may arise in some cases, the slope technique gives a reliable first approximation to the pitch angle values. The steps we followed to apply the slope technique in our multiple-armed spiral galaxies, can be summarized as follows: 1. On an image of the galaxy, we overlay two ellipses centred on the centre of the galaxy; the inner ellipse marks the corotation of the bar, while the outer ellipse shows the projected radial position of the bar corotation plus at least twice its uncertainty (for some galaxies it is possible to extend this ellipse further out as the pitch angle of the spiral arms remains constant over this extended range). 2. We click repeatedly along the arm structure and within the region limited by the two ellipses with the cursor. Doing this we produce a list of projected coordinates \((x_i,y_i)\) of the pixels that define the spiral arm. 3. The coordinates are deprojected and then transformed to polar coordinates, so we finally produce a list of \((\theta_i,\ln{r}_i)\) points. 4. These points are plotted in the \((\theta,\ln{r})\) plane, and a linear fit is performed; according to equation~\ref{eq:pitch}, the pitch angle is calculated as the $\arctan$ of the slope of the linear fit, and the associated uncertainty is obtained from the error on the slope. Most galaxies are two-armed in the radial spiral segment where we apply the Slope method, so we obtain two values of the pitch angle. In this case we give the mean value and uncertainties are calculated accordingly. In some specific galaxies, only one arm is clearly visible so a single pitch angle is determined. These values are listed in column 10 of Table~\ref{tab:table2}.

With the Font-Beckman method we determined the resonant structure of a disc galaxy in terms of corotation radii or pattern speed as each density wave rotating with its own pattern speed occurs in a distinct  annular zone around a specific radius (its corotation radius). Among all peaks found in the histogram of phase-reversals for a given galaxy (see subsection~\ref{sec:bar} for a brief description of how to obtain this histogram, or \citet{Font2014a} for a detailed description), the strongest peak beyond the bar corotation radius, which is assumed to correspond to the dominant density wave, is associated with the corotation of the inner section of the spiral arms. In general, we find more than one corotation resonance for the spiral arm structure (in some galaxies, we measure up to five resonances in the spiral arm region), which is in agreement with the simulations of \citet{Roskar2012}, who found multiple pattern speeds for the dominant $m=2$ amplitude in the disc with a set of models.

\section{results and discussion}

\subsection{Variation of parameters with the morphological type of the galaxy}
In this section we show the variation of the disc mass fraction, the relative angular momentum of the bar, the pitch angle and the difference between the relative pattern speed of the bar and the spiral arms, with the morphological type of the galaxy given in the RC3 catalogue \citep{Vaucouleurs1995}.

\begin{figure*}
	\includegraphics[width=\textwidth]{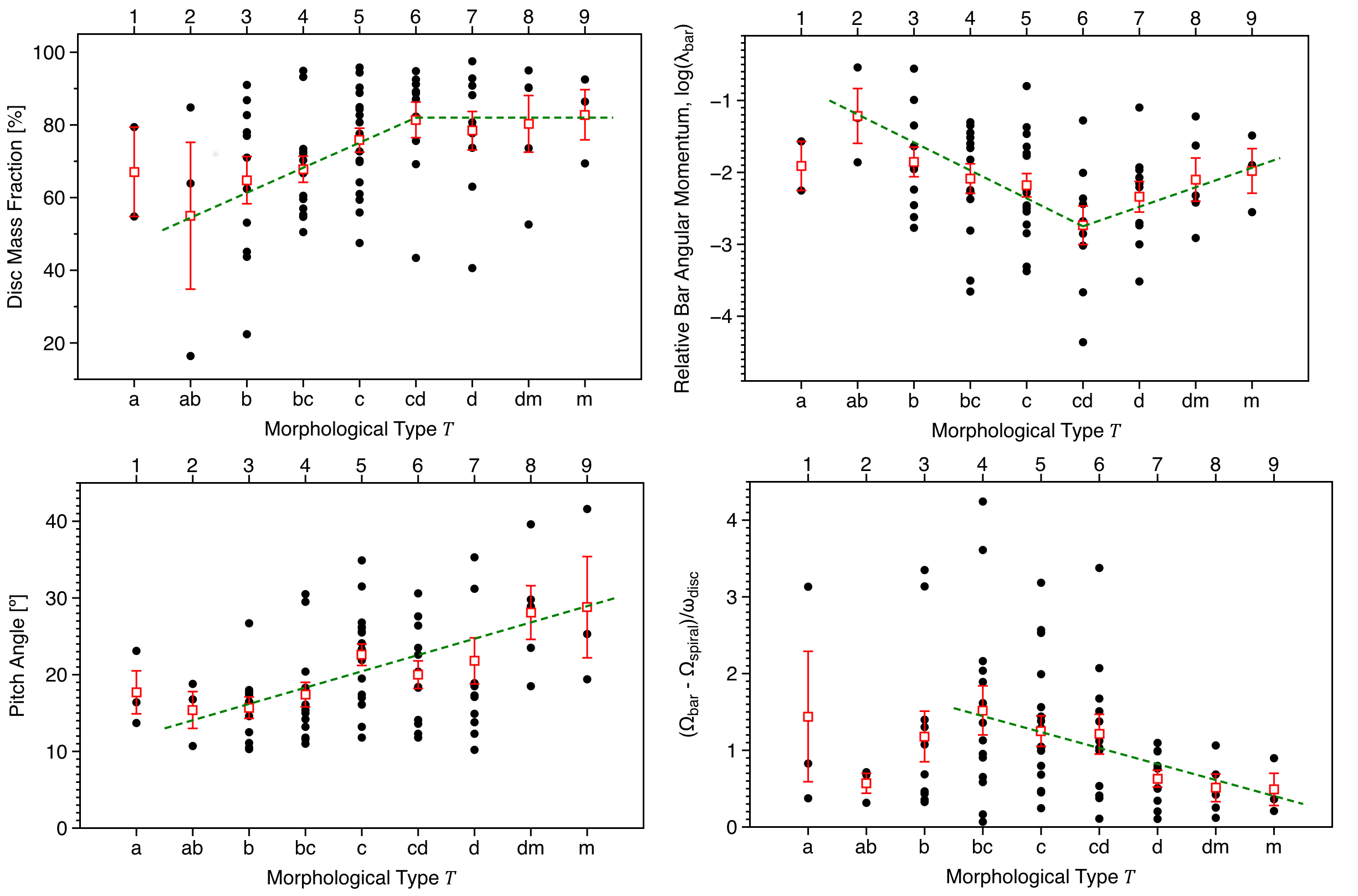}
    \caption{Distribution of the stellar disc mass fraction (top left panel), the relative angular momentum of the bar, $\lambda_{bar}$ (in logarithm, top right panel), the pitch angle of the spiral arms (bottom left panel) and the difference between the pattern speeds of the bar and the spiral arms relative to the disc angular velocity, $\delta_{B,S}$ (bottom right panel) as a function of the morphological type of the galaxy given in the RC3 catalogue. In all panels, the red boxes mark the mean value of the corresponding parameter for each morphological type, and the error bars show the error on the mean. The dashed lines in green are overplotted to highlight the behaviour of the parameter along the Hubble sequence; in the two top panels, the green lines correspond to the linear fit of the average values for two intervals of morphological type (excluding those few galaxies of type $T=1$), whereas for the pitch angle a single linear fit is performed, and for the $\delta_{B,S}$ parameter only intermediate and late type galaxies are taken into account.}
    \label{fig:Hubble1}
\end{figure*}

The distribution of the mass fraction of the disc with the galaxy Hubble type is plotted in the top-left panel of Fig.~\ref{fig:Hubble1}, in which the red boxes show the mean of the disc mass fraction for each morphological type, and the error bars depict the error of the mean. There is a significant decline in the value of this parameter from SBa-type galaxies to SBab-type, (although the numbers for the two types of galaxies are quite small), and the Figure then shows that the disc mass fraction tends to increase monotonically from the lowest values until SBcd-type galaxies ($T=6$), while, for later type galaxies, $6\leqslant T\leqslant9$, the average mass of the disc relative to the total mass of the galaxy remains nearly constant. This behaviour is highlighted by means of the dotted line in green.

\begin{figure*}
	\includegraphics[width=\textwidth]{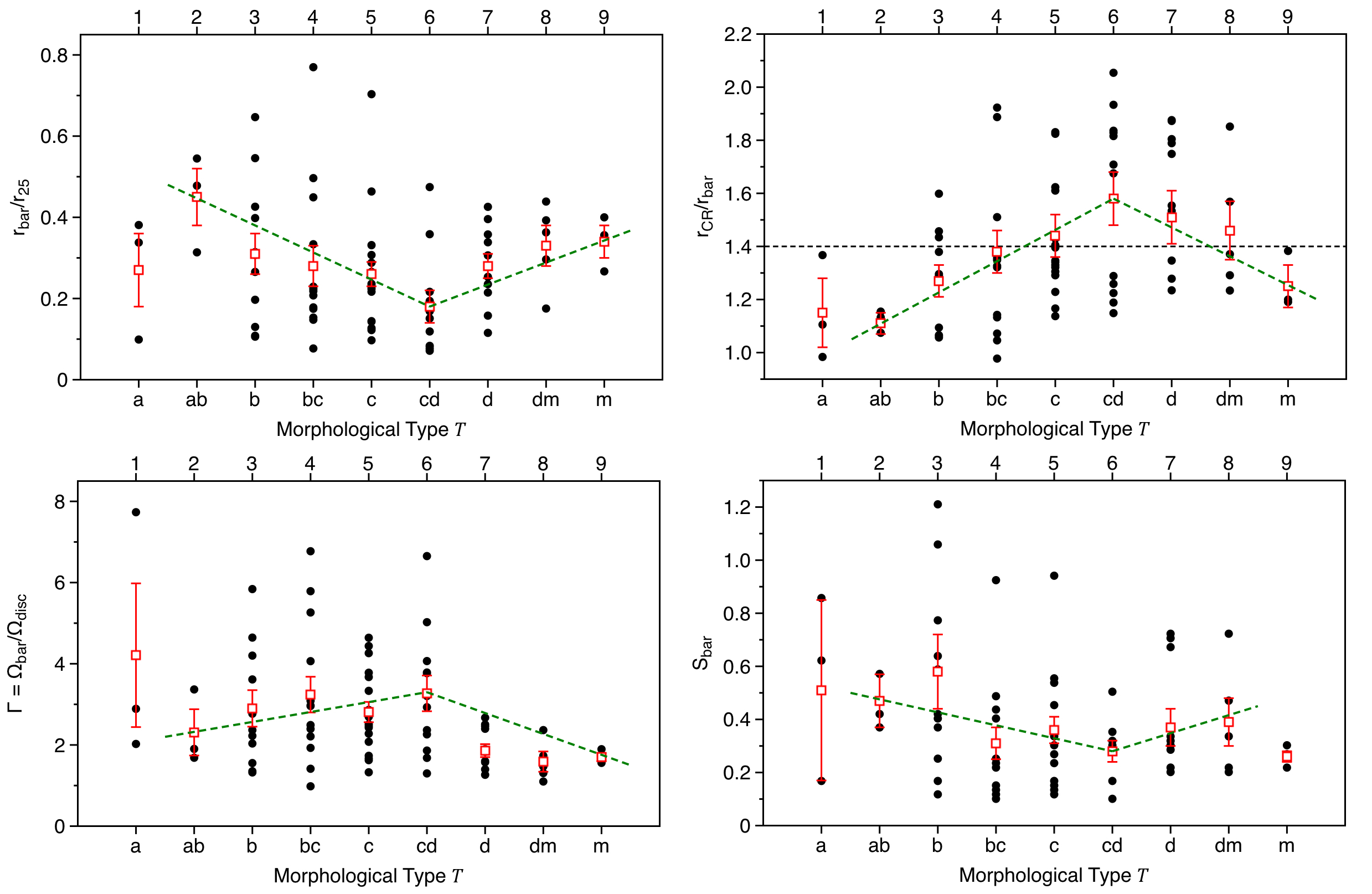}
    \caption{Variation of the relative bar length (top left panel), the rotational parameter (top right panel), the bar strength (bottom left panel) and the relative pattern speed of the bar (bottom right panel) as a function of the morphological type of the galaxy. In all panels, the red boxes mark the mean value of the corresponding parameter for each morphological type, and the error bars show the error of the mean. The dashed lines in green are the linear fits of the average values for two intervals of morphological type, showing the behaviour of this parameter along the Hubble sequence. The dashed horizontal line, in black, in the top-right panel marks the separation between slow/fast rotators as conventionally defined.}
    \label{fig:Hubble2}
\end{figure*}

Fig.~\ref{fig:Hubble1}, top-right panel, illustrates the variation of the scaled bar angular momentum, $\lambda_{bar}$, with the galaxy morphological type. This parameter shows a large rise from SBa-type galaxies to SBab-type, although this could be an effect of the poor statistics of galaxies of a-type in our sample. We can see that bars in earlier type galaxies (but with $T\ne1$) show larger values of the relative angular momentum than those measured for bars hosted by galaxies of type $T=6$, for which this parameter reaches its minimum value; then the tendency is reversed and the relative angular momentum of the bar increases through the galaxies of later type. This feature is highlighted in the plot using green dashed lines which are calculated as the linear fits of the average values for two intervals of morphological type (excluding galaxies of type $T=1$). It is important to point out that other fundamental parameters that characterize the bar also show a "break" in their distribution along the Hubble sequence for galaxies of type $T=6$, which is illustrated in Fig.~\ref{fig:Hubble2} where the variation of four bar parameters is plotted along the Hubble sequence. This figure shows that the relative bar length (in the top left panel) has a local minimum for SBcd galaxies; this feature was also found by \citet{Martin1995, Laurikainen2007, Font2017}. The relative bar pattern speed also shows a local minimum at $T=6$, while the distribution of the bar strength and the ratio of the bar corotation radius to the bar length (the so-called rotational parameter) reach a local maximum for galaxies of type $T=6$. In addition, we also found a change of the behaviour through the Hubble sequence of the disc mass fraction for galaxies starting at SBcd-type ($T=6$ and above). 

The variation of the pitch angle of the spiral arms along the Hubble sequence is plotted in the bottom-left panel of Fig.~\ref{fig:Hubble1}. After an initial decrease of the pitch angle from galaxies of type $T=1$ to galaxies of type $T=2$, the pitch angle shows a tendency to increase from earlier type galaxies to later type. In other words, spiral arms are tightly wound in galaxies of earlier type, while galaxies of later type show more open arms. The feature is clearly shown by the green dashed line in the figure, which is obtained as a linear fit of the averaged values of the pitch angle as a function of the morphological type. The slope of the linear fit has a value of $\sim$2, which is in agreement with the correlation suggested by \citet{Roberts1975}, based on the predictions of the Density Wave theory of \citet{Lin1964}; according to this theory, galaxies with less dominant discs should develop tighter spiral arms; this is what can be inferred from the distribution of the disc mass fraction and the pitch angle (top-right and bottom-left panels of Fig.~\ref{fig:Hubble1}, respectively): late type galaxies have more dominant discs and the spirals arms are less tightly wound, while discs in earlier galaxies are less massive with tightly wound arms. A similar tendency to grow of the pitch angle along the Hubble sequence between SBa and SBc galaxies is also reproduced by \citet{Kennicutt1981} with a sample of 113 objects, although \citet{Seigar1998} found an almost uniform distribution of the pitch angle in the Hubble sequence with a sample of 45 face-on galaxies. We define the dimensionless $\delta_{B,S}$ parameter as the difference in pattern speed between the bar and the spiral arms, relative to the angular velocity of the disc. The variation of this parameter along the Hubble sequence is plotted in the bottom-right panel of Fig.~\ref{fig:Hubble1}. This parameter characterizes the shear rate between the central bar and the arm structure of the galaxy, and intuitively it should be anti-correlated with the pitch angle of the arms: loosely wound arms should be found in galaxies where the bar pattern speed is similar to the spiral pattern speed, while arms that rotate significantly slower than the bar should be tightly wound. This relationship is qualitatively reproduced in the distribution of the pitch angle and the $\delta_{B,S}$ parameter (see the two bottom panels of Fig.~\ref{fig:Hubble1}) only for galaxies of intermediate and late type ($4\leqslant T\leqslant9$), for which the pitch angle tends to increase while $\delta_{B,S}$ shows a tendency to decrease. The poor statistics of earlier type galaxies could be the reason why these galaxies do not, apparently, follow the anti-correlation relationship found for intermediate and later type galaxies.

\subsection{Interplay between the arms and the bar}

It is well established that bars can induce a spiral structure in disc galaxies (see \citet{Dobbs2014} for a general review). The problem was initially tackled theoretically by \citet{Oki1964,Linden1979,Roberts1979}, and with analytical methods by \citet{Lin1975}. All this early work had counterparts in hydrodynamic numerical simulations \citep{Sanders1976,Sanders1977,Huntley1978, Berman1979, Athanassoula1980, Schempp1982,Wada1994,Dobbs2010,Baba2015,Sormani2015}, which modelled the mechanism of spiral arm formation by the effect of the  bar. On the observational side, mechanisms for spiral density waves were studied by \citet{Kormendy1979} with a sample of 25 barred galaxies and 8 galaxies with a companion. \citet{Elmegreen1985} investigated the properties of bars and spirals with a sample of 15 barred spiral galaxies of different morphological types. In \citet{Seigar1998} a sample of 45 face-on spiral galaxies was analysed and the results were used to discuss the validity of several version of spiral density wave theory. \citet{Block2004} and later \citet{Buta2009} studied the bar-driven spiral arm mechanism with a sample of 15 and 23 barred spiral galaxies observed in the near-infrared $K_s$ band, respectively. 

We have measured here several parameters characterizing the bar, such as its length, the pattern speed, the angular momentum and the bar strength. We have also determined some relevant properties of the arms of the galaxies: the pattern speed of the spiral structure and the pitch angle of the spiral arms. We study the influence of the bar on the spiral structure by interpreting the relationship between these two sets of parameters.

It is interesting to note, from Table~\ref{tab:table2}, that in general the pattern speeds of the bar and the first spiral arm segments are different. We should first explain that the corotation of the bar is, in virtually all the objects, at a radius greater than the bar length, a condition found quite generally in previous publications \citep{Contopoulos1980,Athanassoula1992}. This implies that the innermost, short, segments of the spiral arms fall within this radius, so that these segments do in fact share the pattern speed of the bar. So when we refer to the ''first spiral arm segments'' we mean the first segments beyond the bar corotation. These segments, then are characterized by their own separate density wave with its own pattern speed. This is in agreement with the results of the numerical simulations of \citet{Sellwood1988, Masset1997, Rautiainen1999, Minchev2012} who all concluded that the bar and the spiral structure should rotate with different pattern speeds. However, in a small subset of seven galaxies with very slowly rotating bars (\(\Omega_{bar} \lesssim 20~kms^{-1}kpc^{-1}\)) the pattern speeds of the first spiral arm segments are very close to those of the bars. We suggest that a reasonable interpretation of these cases is that they correspond to an evolutionary stage of the galaxy in which the bar has accumulated considerable mass and has been braked by the surrounding galaxy, both halo and disc components. This evolutionary braking process which accompanies mass accretion by the bar has been modelled by a number of authors \citep{Athanassoula2003,Martinez2006,Font2017}. The braking is more efficient for the bar than for the spiral arms, so that in a galaxy with a well evolved, massive bar, we would expect its pattern speed not to be much greater than that of the first arm segments. The material in Table~\ref{tab:table2} also contains interesting information about the dynamical coupling between bars and spiral arms, which merits careful consideration, and will be dealt with in a separate article. 

\begin{figure}
	\includegraphics[width=\columnwidth]{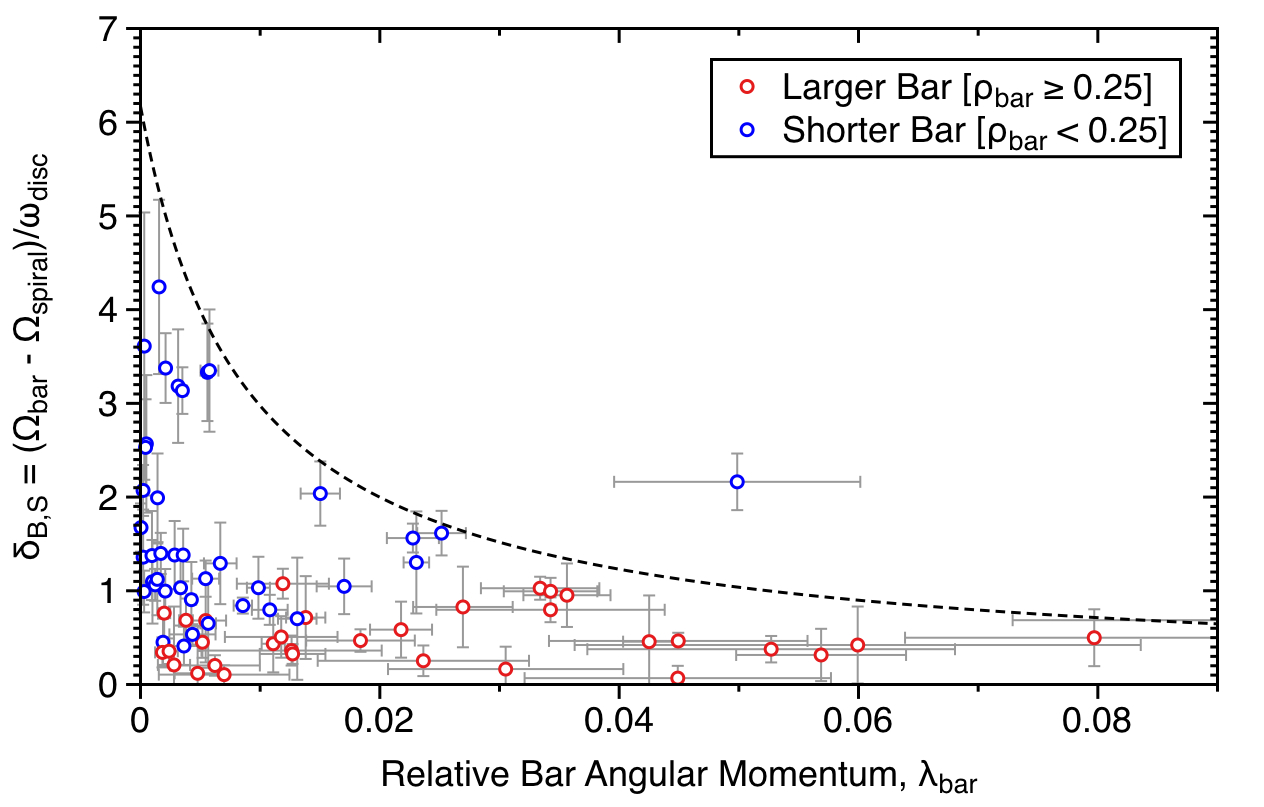}
    \caption{Difference between the pattern speeds of the bar and of the spiral arms divided by the disc angular velocity, $\delta_{B,S}$, as a function of the relative angular momentum of the bar, $\lambda_{bar}$. The dashed line shows qualitatively the envelope of the points in this parametric plane. The data is colour-coded according to the relative bar size.}
    \label{fig:bar_spiral}
\end{figure}

Fig.~\ref{fig:bar_spiral} shows the variation of the scaled angular momentum of the bar, $\lambda_{bar}$, with the difference between the pattern speed between the bar and the arms, $\delta_{B,S}$, which is a measure of the shear between the two structures. We do not find a correlation between the two parameters, although a certain relationship between them is inferred from the figure, which shows five main features: 1. All galaxies (with the exception of UGC9753) in this parametric plane are confined within a well-defined region, which is bounded by an envelope curve of the form $(a+b~x^c)^{-1}$ (plotted as a dashed curve in the figure). 2. Those galaxies hosting a bar with larger angular momentum ($\lambda_{bar} \gtrsim 0.04$), have bars rotating slowly with pattern speeds similar to those of the spiral arms ($\delta_{B,S} \lesssim 0.5$). 3. Those galaxy bars that rotate much faster than the spiral arms ($\delta_{B,S} \gtrsim 3.0$), always have low values of the angular momentum. 4. We do not find any galaxy having a bar with large angular momentum that rotates much faster than the spiral structure. 5. The relative bar size of the spiral galaxies organizes well the data in the (\(\lambda_{bar},\delta_{B,S}\)) plane, where the border between shorter and larger bars (colored in blue and red in Fig.~\ref{fig:bar_spiral}, respectively) can be traced with a simple line of positive slope. The shorter bars ($\rho_{bar} < 0.25$) have low values of angular momentum ($\lambda_{bar} \lesssim 0.03$) and rotate clearly faster than the inner segment of the spiral arms, \(\Omega_{bar} \gtrsim \Omega_{spiral}+\omega_{disc}\); while the larger bars ($\rho_{bar} \geq 0.25$) cover the whole range of the values of the relative angular momentum with the lowest values of the shear parameter ($\delta_{B,S} \lesssim 1$).

\subsubsection{Variation of pitch angle with properties of the galaxy}

In this section we analyse how the pitch angle behaves when is related with other parameters of the galaxy that we have measured. Although no direct correlation is found between the pitch angle and any of the parameters considered here, it is possible to set some restrictions for these parameters, which have implications on the different theories of spiral evolution.

\begin{figure}
	\includegraphics[width=\columnwidth]{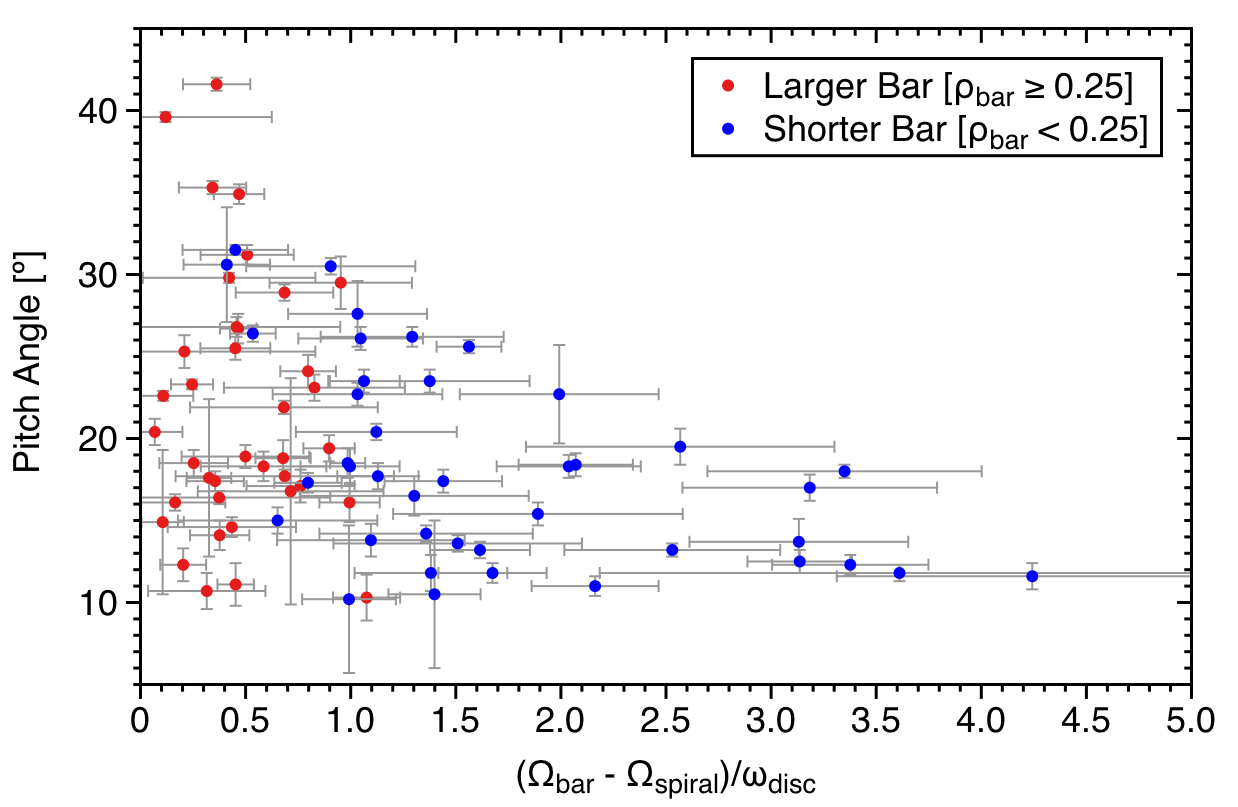}
    \caption{Pitch angle versus the difference between the pattern speeds of the bar and the arms, relative to the disc angular rate, $\delta_{B,S}$. Those galaxies having a bar larger than 25$\%$ of the 25 B-band mag\,arcsec$^{-2}$ radius, r$_{25}$, are plotted in red, while in blue are plotted those galaxies hosting a bar shorter than 25$\%$ of r$_{25}$.}
    \label{fig:pitch_omegas}
\end{figure}

The first parameter to confront with the pitch angle, is the shear parameter, $\delta_{B,S}$, which gives a measure of how much faster the bar rotates with respect to the spiral arms. This is shown in Fig.~\ref{fig:pitch_omegas}, in which we distinguish between galaxies with bars shorter than 25$\%$ of $r_{25}$, and galaxies with larger bars, plotted in blue and red, respectively. The figure shows two main features; firstly the relative bar size, $\rho_{bar}$, organizes quite well the data, so that longer bars are found in the region of low values of the shear parameter, \(\delta_{B,S}\lesssim 1\), while the shorter bars rotate much faster than the spiral arms (\(\Omega_{bar}\gtrsim \Omega_{spiral}+\omega_{disc}\)). This may be expected, since a longer bar will have a greater influence on the outer parts and therefore govern the spiral kinematics with its pattern speed, while a shorter bar will influence the spiral structure less, leaving it free to have a different pattern speed. Note that the classification between shorter/longer bars is somewhat arbitrary, however with the criterion assumed here, the border that splits off these two regimes is a vertical line placed at \(\delta_{B,S}\simeq 1\); this points to an implicit anti-correlation between the shear parameter and the relative bar length. Secondly, all points are uniformly distributed within one half of the parametric plane, having a well-defined linear envelope with negative slope, which limits the region where all galaxies are found. There is no galaxy found outside this parametric region, i.e. we do not find a single spiral galaxy in our sample with open arms that rotate much more slowly than the bar. In other words, the linear envelope also limits a forbidden region in the \((\delta_{B,S},\varphi)\) plane. This figure also reveals that those barred spiral galaxies with more open arms, must harbour a large bar that rotates with a pattern speed similar to that of the spiral arms, while those bars rotating much faster than the spiral arms, can only be developed in galaxies with tighter wound spirals. Additionally, these bars are small in length (\(\rho_{bar}<0.25\)) and must have a low relative angular momentum (according to Fig.~\ref{fig:bar_spiral}) . These results confirm what intuitively one should expect: if a galaxy has a bar which is rotating much faster than the spiral structure, then the spiral arms should be tighter wound, and if a barred galaxy has a open spiral arms then the bar rotates with a pattern speed slightly larger than that of the spirals. 

\begin{figure}
	\includegraphics[width=\columnwidth]{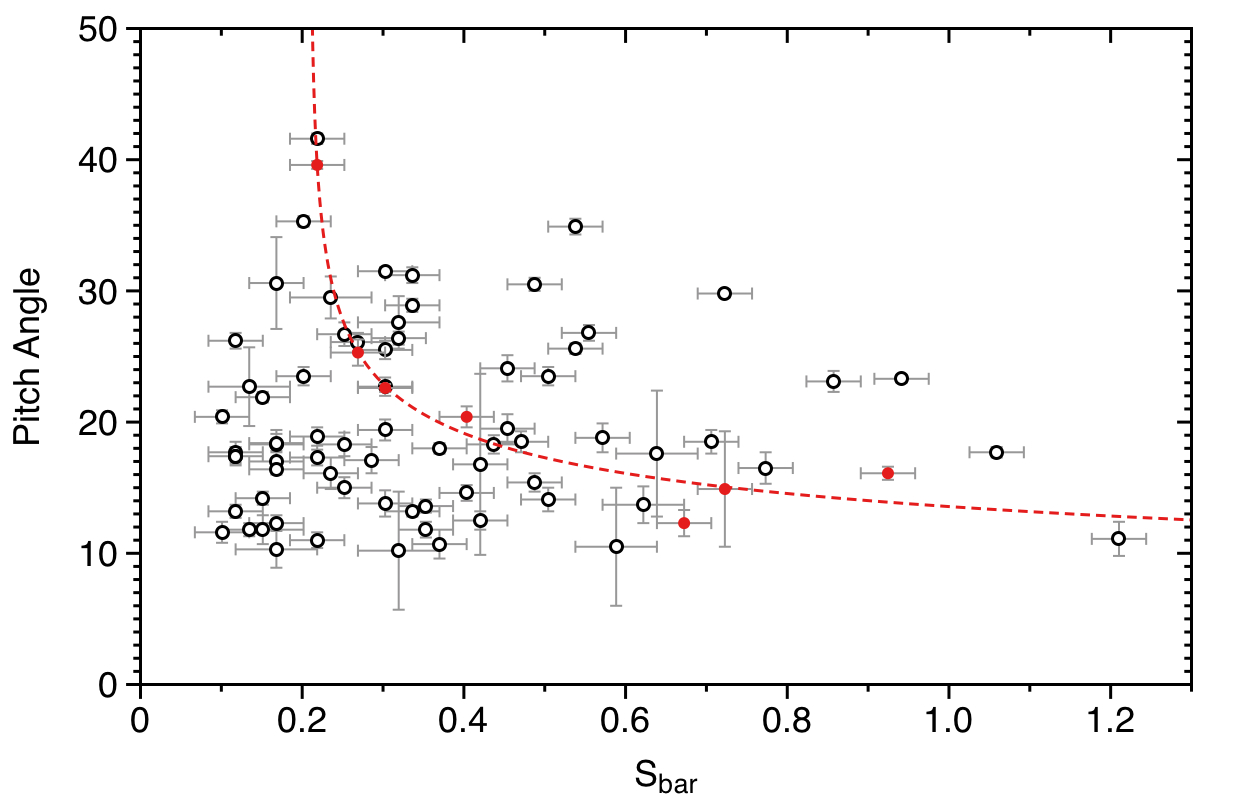}
    \caption{Variation of the pitch angle as a function of the bar strength. Those galaxies in which the bar pattern speed is  close to that of the spiral arms are plotted in red, and the dashed curve in red indicates the fit to these points.}
    \label{fig:pitch_strength}
\end{figure}

In Fig.~\ref{fig:pitch_strength} we display the variation of the pitch angle versus the bar strength. As with other parameters, the measured pitch angle and the bar strength are confined within a well defined region, which is limited by a linear envelope; this shows that spiral galaxies with a stronger bar can only have a spiral structure with tightly wound arms, and also that an open spiral is found only in galaxies with a weaker bar. As shown in this figure, we do not find a galaxy in our sample that has a strong bar with open arms. This result does not favour the bar-driven spiral invariant manifold theory, which predicts that stronger bars should have less tightly wound arms (i.e. larger values of the pitch angle) than weaker bars \citep{Voglis2006,Romero-Gomez2006,Romero-Gomez2007,Tsoutsis2008,Tsoutsis2009,Athanassoula2009, Athanassoula2010,Athanassoula2012}; \citet{Martinez2012} analysed a sample of 27 galaxies and found that only a small subset of seven galaxies seems to corroborate this trend, however the author used Fourier methods to determine the pitch angle as well as the bar strength. This disagreement persists even if we consider only those galaxies in which the bar and the spirals rotate with closely similar values for their pattern speeds (\(\Omega_{spiral}\simeq \Omega_{bar}\)), which are easily identified from Table~\ref{tab:table2}, columns (7) and (9). This yielded a subset of seven galaxies, which are colored in red in Fig.~\ref{fig:pitch_strength}, these galaxies are showing a strong correlation between the pitch angle and the bar strength, which are well fitted with a function of the form $(a~(x-b)^c)^{-1}$ (plotted as dashed red line), with a correlation coefficient $r^2 = 0.97$. The interpretation of this correlation is potentially interesting and maybe numerical simulations would be needed to verify it, but this is beyond the scope of this study. 

\begin{figure}
	\includegraphics[width=\columnwidth]{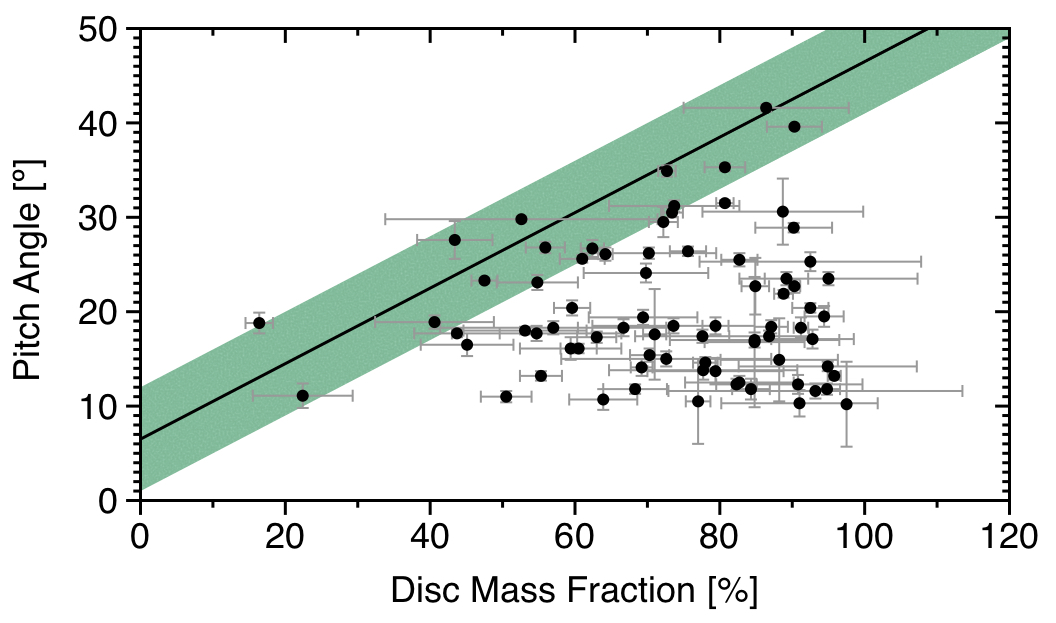}
    \caption{Pitch angle of the inner spiral arms plotted against the stellar disc mass fraction. Solid line marks the envelope of the points in this parametric plane; the envelope 2$\sigma$ region (shadowed in green) is also displayed.}
    \label{fig:pitch_disc}
\end{figure}

The variation of the pitch angle of the spiral arms with disc mass fraction of the baryonic component is shown in Fig.~\ref{fig:pitch_disc}.  All the points are below a linear envelope line with positive slope value, which is plotted in the figure as a solid line, and the 2$\sigma$ region around the envelope line is displayed as the shadowed region in green. This graph reveals the following constraints: 1. There are no galaxies having spirals with open arms and low values of the mass fraction of the disc (i.e. high values of the bar mass fraction). 2. A galaxy with a low value of the relative mass of the disc must have tightly wound spiral arms. 3. If the spiral arms are loosely wound, then the disc must contain most of the mass of the galaxy. Note that this latter constraint does not mean that if the spiral arms are open, then all the mass of the galaxy must be in the disc, as in our sample the galaxies with massive disc have spiral arms with a very wide range of pitch angles.

\begin{figure}
	\includegraphics[width=\columnwidth]{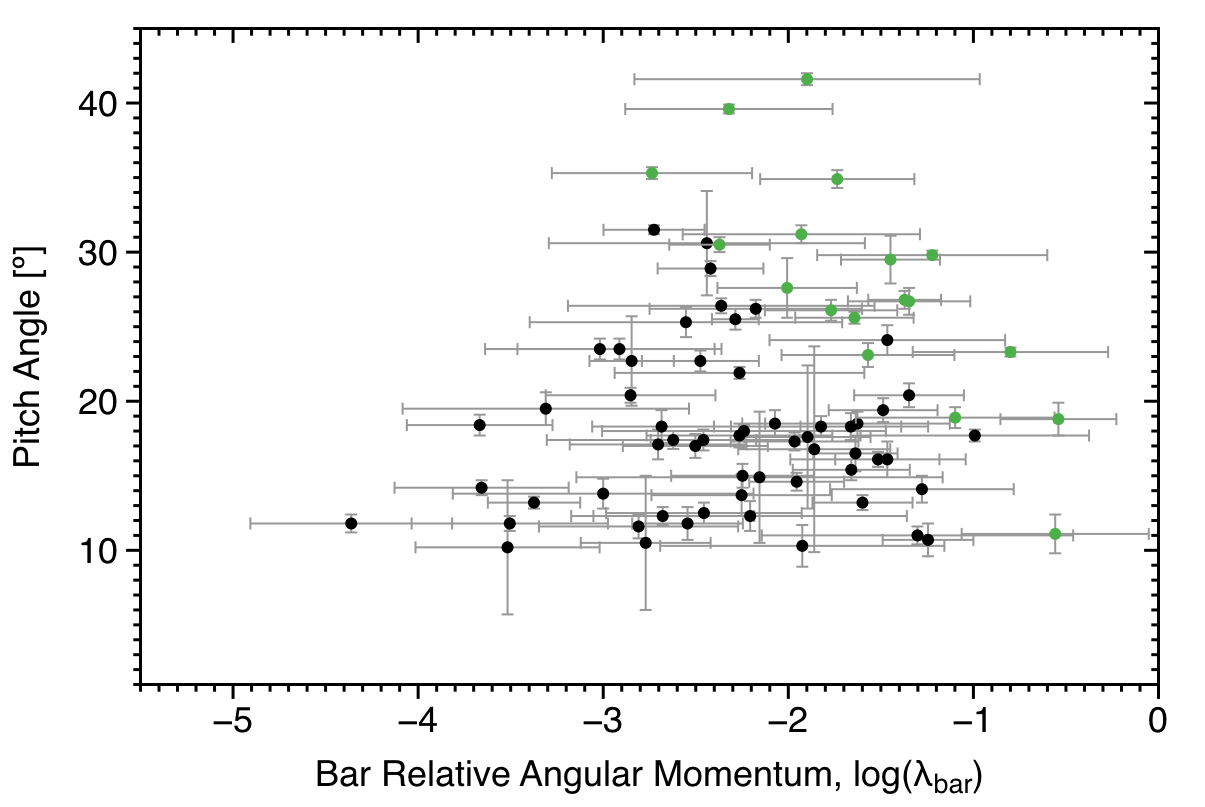}
    \caption{Pitch angle versus the logarithmic relative angular momentum of the bar, $\lambda_{bar}$. In green we show those galaxies that fall within the envelope region of Figure~\ref{fig:pitch_disc}}
    \label{fig:pitch_moment}
\end{figure}

In Fig.~\ref{fig:pitch_moment} the pitch angle of the spiral structure measured in the region dominated by the bar corotation is plotted against the relative angular momentum of the bar, $\lambda_{bar}$ (on a logarithmic scale). Examining the region where the spiral galaxies are distributed in this parametric plane, we notice that all points are bounded by two different envelopes: a first linear envelope with positive slope, which limits those galaxies that host a rotating bar with lower angular momentum, and a second also linear envelope but with negative slope value for the bars with higher angular momentum; the crossing point between the two envelope lines is at $\lambda_{bar} \simeq 10^{-2}$, which means that the most wide open spiral arms are found in galaxies which have a bar with a relative angular momentum of $10^{-2}$. It is interesting to note that if the bar has either a very low value of the relative angular momentum or has a rather large value, then the spiral arms are tightly wound. We identified 15 galaxies that fall within the envelope region in the \((\mu_{disc},\varphi)\) plane (shadowed region in Fig.~\ref{fig:pitch_disc}). This subset of galaxies is plotted in green in the parametric plane, \((\lambda_{bar},\varphi)\) of Fig.~\ref{fig:pitch_moment}, in which we can see that they are distributed within a region defined by the second envelope of negative slope. This means that given a pitch angle of the spiral arms, those galaxies for which the disc mass fraction is minimum also host a bar with maximal relative angular momentum, and \textit{vice versa}. This behaviour of the pitch angle, the mass of the disc and the angular momentum of the bar, outlined in these two figures, reveals that these three parameters are not independent, at least for a specific type of galaxies. 

\subsection{Angular momentum transfer of the bar}

The evolution of bars in disc galaxies is governed by the redistribution of the angular momentum between the different structures in play: the bar, the disc and the halo. This mechanism has been specifically studied in numerical simulations for pure stellar discs \citep{Martinez2006,Villa2009,Collier2018}, and also including the gas component in the disc \citep{Berentzen2007,Villa2010}. The exchange of the angular momentum of the bar with the halo and the outer disc is responsible for the growth in bar size and the slowdown of the rotating bar, as shown in a large number of numerical simulations \citep[see the general review by][and references therein]{Athanassoula2013}. The bar resonances (corotation, inner and outer Lindblad resonance, inner and outer ultra-harmonic 4:1 resonance) play a key role in this process, as the transfer of angular momentum occurs in regions defined by these resonances \citep{Athanassoula2003,Martinez2006,Villa2009}; for example, Fig. 10 of \citet{Villa2009} illustrates, for one of the models used, the regions of the disc and the halo where the angular momentum is emitted or absorbed. Alternatively, the dynamical friction offers a complementary framework which describes bar evolution in disc galaxies, according to the bar-halo friction mechanism, whereby a rotating bar in a halo grows in size as it is braked by friction \citep[see the general review by][and references therein]{Sellwood2014}. 

\begin{figure}
	\includegraphics[width=\columnwidth]{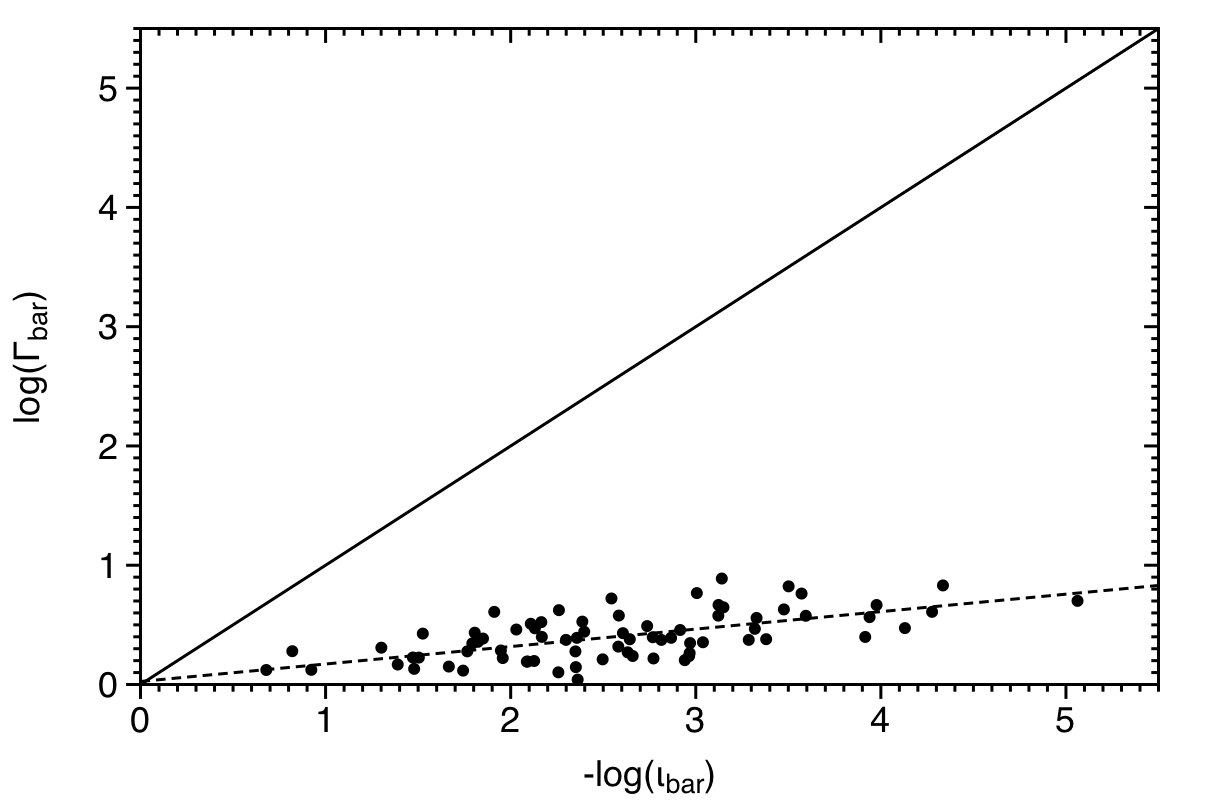}
    \caption{Pattern speed associated with the bar relative to the disc angular velocity, $\Gamma_{bar}$ in logarithm, versus the relative moment of inertia of the bar, $\iota_{bar}$ also in logarithm (note that the sign is reversed). The solid line marks the 1:1 relation and the dashed line is the linear fit to the points. The uncertainty bars are not plotted to better illustrate the correlation between the two bar parameters.}
    \label{fig:slowdown}
\end{figure}

In numerical simulations it is possible to play a movie of the bar evolution in disc galaxies with a dark matter halo, from the formation of the bar until its death. On the observational side, we only have an instantaneous snapshot of that movie as it is more complicated to resolve galaxies well beyond the local Universe. However, not all galaxies in that snapshot are at the same stage of evolution, so we can take advantage of this to analyse the galaxies, and hence, to infer the signatures of the bar evolution. To do so, it is essential to perform precise measurements of dynamical properties of these barred galaxies, such as the pattern speed of the different density waves coexisting in the galaxy. The lack of this type of measurement accounts for the scarcity of observational studies on bar evolution. In that sense, in \citet{Font2017}, we showed how the relation between the measured bar pattern speed and the bar strength supports the results of numerical simulations of the bar evolution. In the present study, in addition to the bar pattern speed, we also calculated the moment of inertia of the bar, $\iota_{bar}$, and the relation between these two bar parameters is plotted in Fig.~\ref{fig:slowdown}. We find a clear anti-correlation between the relative moment of inertia of the bar and the bar pattern speed relative to the disc angular velocity (note that the sign of $\iota_{bar}$ in the figure,in logarithmic scale, is reversed), showing that bars with a large moment of inertia are rotating slowly, whereas fast rotating bars have loosely moments of inertia. The dashed line corresponds to the linear fit to the points. This result can be interpreted in terms of bar evolution; as the galaxy evolves, the angular momentum afforded by the bar is consumed by the outer disc and the halo. This mechanism has three main consequences for the bar: (i) the bar rotates slower, reducing its pattern speed. (ii) The bar becomes longer and/or more elongated. (iii) The bar increases its mass. The final two points imply that the moment of inertia of the bar increases. In conclusion, galaxies with a slower rotating bar are in a more advanced stage of evolution and hence should have larger values of the bar moment of inertia than those bars less evolved which are spinning at higher angular speed and have lower moment of inertia, which is in agreement with our results plotted in Fig.~\ref{fig:slowdown}.

The linear dependency between the two quantities illustrated in Fig.~\ref{fig:slowdown} can be used as a proof of the loss of angular momentum of the bar in spiral galaxies; writing the relative momentum of the bar, defined by equation~\ref{eq:angular}, as the bar angular momentum, $J_{bar}$ divided by the outer disc angular momentum, $J_{disc}$, and taking time derivatives, we have

\begin{equation}
    \dot{\lambda}_{bar}=\frac{\mathrm d}{\mathrm d t} \bigg( \frac{J_{bar}}{J_{disc}} \bigg)= \bigg( \frac{1}{J_{disc}} \bigg) \dot{J}_{bar} - \bigg( \frac{J_{bar}}{J_{disc}^2} \bigg) \dot{J}_{disc}
	\label{eq:derivative}
\end{equation}

the angular momentum that the bar emits, thus \(\dot{J}_{bar}<0\), is employed to feed the halo and the outer disc, so \(\dot{J}_{disc}>0\). Taking these inequalities into account in the equation~\ref{eq:derivative}, we obtain that \(\dot{\lambda}_{bar}<0\), which means that the bar loses relative angular momentum while is evolving. Knowing this, and taking time derivatives to $\lambda_{bar}$ from equation~\ref{eq:angular}, we have

\begin{equation}
 \dot{\lambda}_{bar} =  \dot{\iota}_{bar} \Gamma_{bar} + \iota_{bar} \dot{\Gamma}_{bar} < 0
 	\label{eq:derivative2}
\end{equation}

 Re-arranging terms in the inequality and integrating over time between $t_0$ and $t$, we obtain

\begin{equation}
    \log(\Gamma_{bar}) < -\log(\iota_{bar}) + \log(\lambda_{bar}^0)
	\label{eq:inequation}
\end{equation}

where $\lambda_{bar}^0 = \Gamma_{bar}^0 \iota_{bar}^0$, is the initial value of the relative angular moment of the bar. In Fig.~\ref{fig:slowdown}, we can see that the above inequality is satisfied qualitatively by all spiral galaxies of our sample, as all points in this parametric plane fall under the 1:1 proportion line (solid line), consequently this sets a constraint to the initial value of the relative angular moment of the bar; $\lambda_{bar}^0\nll 1 $. This becomes clearer when we perform a linear fit of the data (dashed line in the figure), so that \( \log(\Gamma_{bar}) = m (-\log(\lambda_{bar}))+n\), obtaining a value for the slope of $m=0.15\pm0.02$ and for the intercept of $n = 0.025\pm0.057$. The intercept value implies that $\lambda_{bar}^0\approx 1 $, and the slope of the linear fit, $m < 1$, confirms numerically that equation~\ref{eq:inequation} is satisfied by all our galaxies. In consequence, it means that the bars of the spiral galaxies are losing angular momentum, and therefore slowing down while growing in size and mass.

\begin{figure}
 	\includegraphics[width=\columnwidth]{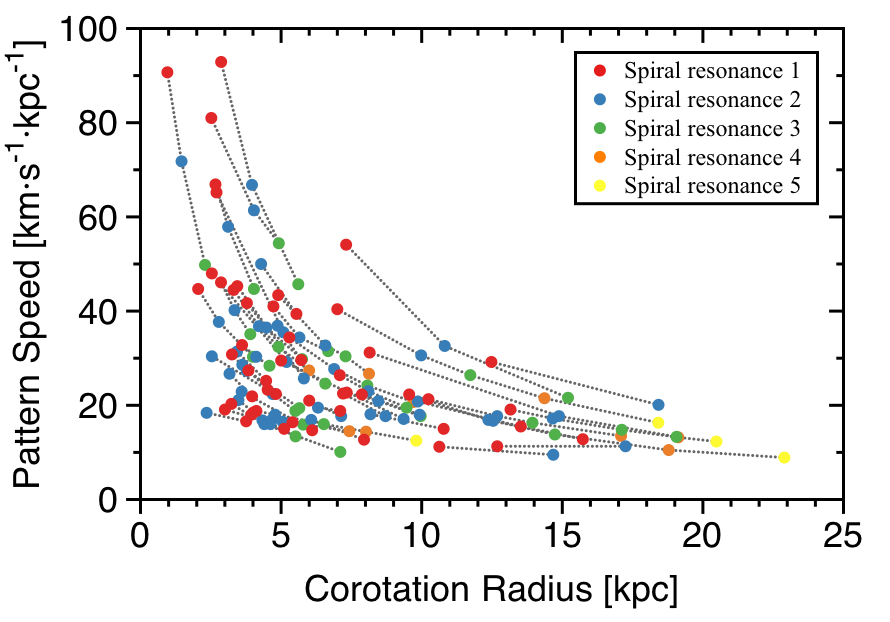}
    \caption{Pattern speed against the corotation radius for those spiral galaxies with at least two resonances propagating in the spiral arm region. The different resonances of each galaxy are connected with dotted lines, in grey.}
    \label{fig:resonances}
\end{figure}

\begin{figure}
 	\includegraphics[width=\columnwidth]{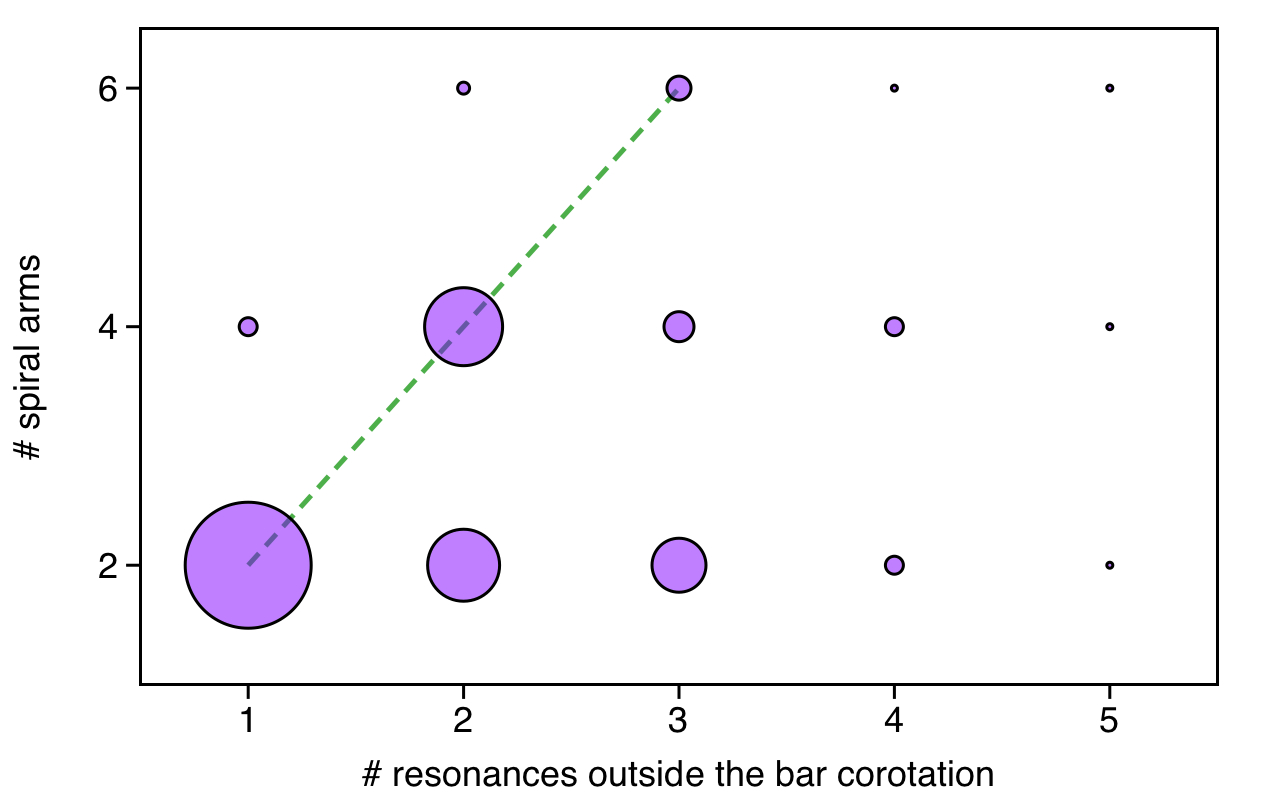}
    \caption{Number of spiral arms, including bifurcations, plotted against the number of the resonances found beyond the bar corotation, which are determined using the Font-Beckman method. The size of each circle is proportional to the number of galaxies that have the specific number of arms and of resonances. The dashed line in green, shows the proportion 2:1 between these parameters.}
    \label{fig:bubbles}
\end{figure}

A further observational result which can be used to probe the different theories relating the dynamical formation of bars and arms is shown in Fig.~\ref{fig:resonances}. Here we have plotted the pattern speed against the corotation radius for those galaxies where we have found at least two resonances propagating concentrically in the region of the disc containing the spiral arms. To show systematic trends we have connected the plotted points for a given galaxy, joining them with straight lines. In the following section we will explain how this result can be used to differentiate different scenarios for arm formation. Fig.~\ref{fig:bubbles} shows a further result of interest  where we plot the number of arms against the number of resonances we determine applying the Font-Beckman method, which fall outside the bar corotation radius. The size of each circle is proportional to the number of the galaxies that have the specific number of arms and resonances. Reading the figure along the rows, we see that between one and five density waves can coexist in grand design galaxies ($m=2$), in which two-armed spiral galaxies with a single pattern speed is the most abundant group, with the remaining groups in descending order until those galaxies with five resonances. The same general behaviour is also found for multi-armed galaxies ($m=4,6$), for which the dominant group are spiral galaxies with four/six arms, which have experienced one/two bifurcations, and harbour two/three density waves in the spiral region.

\section{Conclusions}
With the aim to find observational evidence for the influence of the bar in the spiral arm structure, we have determined the resonant structure of a sample of 79 disc galaxies (listed in Table~\ref{tab:table1}) using high resolution velocity fields. To be specific we have performed accurate measurements of the corotation radius and the pattern speed of the bar and the spiral arms for each object, but other corotations outside the bar region are also found. In addition we measured geometrical parameters using near-infrared images: the bar length and the pitch angle of the spiral arms. Combining these measurements, we could estimate the angular momentum of the bar relative to the outer disc, $\lambda_{bar}$, and the shear parameter, $\delta_{B,S}$. Comparing bar parameters with spiral parameters, we do not find any simple correlation, but all the disc galaxies are distributed within a region of the parametric plane, which is well-defined with a simple envelope (a linear envelope for most of the parameters).This implies that there are forbidden regions in the parametric plane where no galaxy is found, which imposes some constraints to the involved parameters.

The observational results of our study, which are presented in the previous section, can be used to test some of the theories of spiral arm formation in barred galaxies \citep[For a concise description of all theories, see the general review by ][]{Dobbs2014}. The quasi-stationary density wave theory (QSDW) developed by \citet{Lin1964} assumes that spirals arms are density waves propagating within the galaxy disc where stars and gas are stacked while these components are passing through these dense regions \citep[see][and references therein, for a detailed treatment]{Shu2016}. According to this theory, galaxies with $m=2$ spiral structure are favoured contrary to disc galaxies with $m\geq3$ spiral arms, which are less likely. With the near-infrared images we have counted the number of spiral arms of the galaxies of our sample by visual inspection, and we  find only galaxies with even numbers of spiral arms, distributed as follows: 56\% grand design galaxies ($m=2$), 34\% with $m=4$, and the remaining 10\% with $m=6$ spiral arms, where the two latter galaxies are essentially grand design galaxies in which each of the two spiral arms presents one or two bifurcations at different radius. These numbers tend to support the predictions of the density wave theory, although bar driven theories also predict this preference for $m=2$ spiral structures \citep{Kormendy1979}. Another prediction of this theory is that the pitch angle should be anti-correlated with the central mass fraction, which implies that it should be correlated with the disc mass fraction; we do not find such a correlation. However, in Fig.~\ref{fig:pitch_disc} we show that the envelope curve in the parametric plane ($\mu_{disc},\varphi$) is linear with a positive slope. This implies three main constraints: (1) There is no barred galaxy with open arms and a low disc mass fraction. (2) Those galaxies with larger central mass fraction (low values of $\mu_{disc}$) have spiral arms which are tightly wound. (3) Those barred galaxies with loosely wound spiral arms are  found only in discs containing most of the mass of the galaxy. We should emphasize that the mass fractions referred here, and throughout the article, are fractions of the stellar mass (i.e. essentially the barionic mass) of the galaxy within the limiting radius of our measurements. A third prediction of the QSDW theory is that pitch angle of spiral arms increases along the Hubble sequence \citep{Roberts1975,Kennicutt1981}. Although the error bars of the mean values shown in the bottom-left panel of Fig.~\ref{fig:Hubble1} are significantly larger for earlier and later type spiral galaxies, we do find a weak correlation between these two parameters. These results are not in full agreement with the quasi-stationary density wave theory, but they tend to favour its predictions. One main assumption of this simplest version of density wave theory is that spiral structure should be rotating rigidly with a single pattern speed.  We find this feature in only 30\% of the observed spiral galaxies, most of them being grand design galaxies ( 88\% of this subset of galaxies), the remaining 70\% show more than one density wave coexisting in the the spiral structure region. From this, we conclude that the quasi-stationary density wave theory in not enough to describe the spiral structure of our sample of disc galaxies, so that other mechanisms, which in principle are not mutually exclusive, must be taken into account.

A different approach to tackle the formation and evolution of spiral arms in barred galaxies, is adopted in the is adopted in the theory of "chaotic" spirals  \citep{Patsis2006,Voglis2006,Romero-Gomez2006,Romero-Gomez2007,Tsoutsis2008,Tsoutsis2009,Athanassoula2009, Athanassoula2010,Athanassoula2012}. This theory is based on the orbital motion determined by the invariant manifolds associated with the unstable Lagrangian points at the corotation region. The manifold theory in its present form applies to galaxies, where the spiral corotates with the bar. The predicted according to this scenario properties of the spiral arms are based on this assumption. However, our analysis shows that the pattern speed of the bar and the innermost independent segment of the spiral arms are in most cases different (cf. columns (7) and (9) of Table~\ref{tab:table2}). We find that the spiral arms rotate more slowly than the bar, as has also been suggested in some of the simulations in the literature \citep{Sellwood1988,Rautiainen1999,Baba2009,Grand2012,Baba2015}. This favors the propagation of non-linear density waves \citep{Tagger1987,Sygnet1988}, although a resonant coupling could not be identified in our study. In the framework of the manifold theory it is also predicted that spiral arms are more open in strongly barred galaxies than for weak bars. In general we do not reproduce this correlation between the bar strength and the pitch angle calculated for the first segments of the spiral arms, as shown in Fig.~\ref{fig:pitch_strength} where the data are uniformly distributed in a very well-defined region of this parametric space (a similar result is also obtained by \citet{Diaz2018}, who, in a forthcoming article, compare the two parameters for large sample of 400 galaxies from the S4G survey, calculating the pitch angle as a weighted averaged value along the spiral arms). The disagreement becomes even more evident, exactly in the cases when we identify in the ($\mu_{disc},\varphi$) diagram those galaxies in which bar and spiral structures rotate with similar pattern speeds (\(\Omega_{bar}\simeq \Omega_{spiral}\)), showing that the points are well fitted with an inverse polynomial function (red dashed curved in Fig.~\ref{fig:pitch_strength}). This points to an anti-correlation between these parameters for galaxies with bars and spiral arms rigidly rotating. In conclusion, our findings do not support the predictions of the invariant manifold theory, at least within the framework it is developed until now. 

A third theory with predictions about the relations between the parameters measured in the present study is the swing amplification theory \citep{Goldreich1965,Julian1966}, which assumes a continuous source of perturbation in order to maintain the dynamical spiral pattern, otherwise the spiral structure would be dissolved in one or two spiral years \citep{Toomre1991}. The first prediction to test is the anti-correlation between the pitch angle and the shear rate, which is found in numerical simulations \citep{Grand2013}, and in observations \citep{Seigar2006} only when the shear rate is calculated at a fixed radius independent of the disc scale length. In this work we do not measure the shear rate but our observations do produce a value of the shear parameter, which is defined as the difference between the pattern speed of the bar and the spiral arms relative to the angular rate of the outer disc. With our data we do not reproduce a clear anti-correlation.However, in Fig.~\ref{fig:pitch_omegas}, where the pitch angle (pitch angle) is plotted against the shear parameter, all the data are uniformly distributed under an linear envelope which does have the predicted anti-correlation.  Another prediction of the swing amplification mechanism is that the number of spiral arms should tend to increase with the radius \citep{Athanassoula1987, Donghia2015}, which can be easily checked by visual inspection of near-infrared images. We found that 34\% of our spiral galaxies show one bifurcation in each spiral arm, and 10\% of them show two bifurcations in each arm at different radii, which are consistent with the predictions. Such an effect has also been found in numerical simulations of gaseous response models \citep{Patsis1994,Patsis1997}. Another general prediction of the swing amplifier mechanism operating within density wave theory is that this mechanism does not need to assume that the spiral arms rotate in strict corotation with the galactic disc \citep{Baba2013}. This implies that more than one density wave can coexist in the region of the spiral structure, each of them rotating with its own pattern speed, being dominant in a specific annular region, so that the larger the radial extend of the annular region the slower is the rotating density wave. This is confirmed by our observations shown in Fig.~\ref{fig:resonances}.

The two latter predictions are qualitatively illustrated in Fig.~\ref{fig:bubbles}, which shows a preferred 2:1 proportion between spiral arms and resonances, marked with the green dotted line in Fig.~\ref{fig:bubbles}, is in overall agreement with the swing amplification theory.

The 2:1 proportion of Fig.~\ref{fig:bubbles} tells us that each pair of bifurcations (one in each arm) is associated with a new density wave, which rotates more slowly and is dominant in the region beyond the bifurcation radius. Then, why does a significant number of spiral galaxies not conform to  the 2:1 proportion?

A visual analysis of the spiral arms in disc galaxies allows us to distinguish between four different patterns which can be found in the morphology of  arms: (i) the spiral arm is characterized by a single value for the pitch angle over its full extent. (ii) The spiral arm bends abruptly at a given position, which is translated into a change of the slope in the \((\theta,\ln{r})\) plane at a given radius, giving two different values of the pitch angle for that arm. In that case, it is convenient to define segments for the spiral arm according to these "breaks" in each of which a single value of the pitch angle is determined. (iii) The pitch angle varies uniformly with the radius, this means that the \((\theta,\ln{r})\) points of the spiral arm are better fitted by a polynomial than by a linear fit. (iv) The spiral arm can experience more than one bifurcation, increasing the number of spiral arms or "branches"; each of them may have its own pitch angle. These features are not mutually exclusive, they come combined in many different sequences, affecting the global morphology of the spiral arms.

All these scenarios indicate that, in general, the spiral arms are not adequately described by a unique logarithmic curve with a constant pitch angle, highlighting the difficulty of giving a single value of the pitch angle for the spiral arms of a galaxy. There are two different options to overcome this situation: 1. To measure the pitch angle in a specific radial region of the spiral structure. In our study we adopted this procedure, measuring the pitch angle in the innermost segment of the spiral arms. 2. To determine the variation of the pitch angle along the spiral arms, and calculate an average value \citep{Seigar2006}. Answering the question which initiated this section, we believe that the "breaks" in the pitch angle can also be associated with a new density wave in the same way that bifurcations can, so that our graph which includes only the bifurcations is incomplete. However, this requires significant  further investigation, which will be performed in a follow-up study of the spiral structure of barred spiral galaxies.

\section*{Acknowledgements}

This research was carried out with funding from project P/308603 of the Instituto de Astrofísica de Canarias. GHaFas is a visitor instrumnet on the William Herschel Telescope which is operated on the island of La Palma by the Isaac Newton Group in the Spanish Observatorio del Roque de los Muchachos of the Instituto de Astrofísica de Canarias.This research has made use of the Fabry-Perot
database, operated at CeSAM/LAM, Marseille, France. This work is based in part on
observations made with the Spitzer Space Telescope, which is operated by the Jet
Propulsion Laboratory, California Institute of Technology under a contract with
NASA. Funding for SDSS-III has been provided by the Alfred P. Sloan Foundation,
the Participating Institutions, the National Science Foundation, and the U.S.
Department of Energy Office of Science. The SDSS-III web site is
http://www.sdss3.org/. SDSS-III is managed by the Astrophysical Research
Consortium for the Participating Institutions of the SDSS-III Collaboration including
the University of Arizona, the Brazilian Participation Group, Brookhaven National
Laboratory, Carnegie Mellon University, University of Florida, the French
Participation Group, the German Participation Group, Harvard University, the
Instituto de Astrofisica de Canarias, the Michigan State/Notre Dame/JINA
Participation Group, Johns Hopkins University, Lawrence Berkeley National
Laboratory, Max Planck Institute for Astrophysics, Max Planck Institute for
Extraterrestrial Physics, New Mexico State University, New York University, Ohio
State University, Pennsylvania State University, University of Portsmouth, Princeton
University, the Spanish Participation Group, University of Tokyo, University of Utah,
Vanderbilt University, University of Virginia, University of Washington, and Yale
University. We thank Peter Erwin for help in defining and measuring the lengths of
galactic bars.




\bibliographystyle{mnras}
\bibliography{jfont_spirals} 

\begin{thebibliography}{}
\makeatletter
\relax
\def\mn@urlcharsother{\let\do\@makeother \do\$\do\&\do\#\do\^\do\_\do\%\do\~}
\def\mn@doi{\begingroup\mn@urlcharsother \@ifnextchar [ {\mn@doi@}
  {\mn@doi@[]}}
\def\mn@doi@[#1]#2{\def\@tempa{#1}\ifx\@tempa\@empty \href
  {http://dx.doi.org/#2} {doi:#2}\else \href {http://dx.doi.org/#2} {#1}\fi
  \endgroup}
\def\mn@eprint#1#2{\mn@eprint@#1:#2::\@nil}
\def\mn@eprint@arXiv#1{\href {http://arxiv.org/abs/#1} {{\tt arXiv:#1}}}
\def\mn@eprint@dblp#1{\href {http://dblp.uni-trier.de/rec/bibtex/#1.xml}
  {dblp:#1}}
\def\mn@eprint@#1:#2:#3:#4\@nil{\def\@tempa {#1}\def\@tempb {#2}\def\@tempc
  {#3}\ifx \@tempc \@empty \let \@tempc \@tempb \let \@tempb \@tempa \fi \ifx
  \@tempb \@empty \def\@tempb {arXiv}\fi \@ifundefined
  {mn@eprint@\@tempb}{\@tempb:\@tempc}{\expandafter \expandafter \csname
  mn@eprint@\@tempb\endcsname \expandafter{\@tempc}}}

\bibitem[\protect\citeauthoryear{Athanassoula}{Athanassoula}{1980}]{Athanassoula1980}
Athanassoula E.,  1980, \aap, 88, 184

\bibitem[\protect\citeauthoryear{Athanassoula}{Athanassoula}{1992}]{Athanassoula1992}
Athanassoula E.,  1992, \mn@doi [\mnras] {10.1093/mnras/259.2.328}, 259, 328

\bibitem[\protect\citeauthoryear{Athanassoula}{Athanassoula}{2003}]{Athanassoula2003}
Athanassoula E.,  2003, \mn@doi [\mnras] {10.1046/j.1365-8711.2003.06473.x},
  341, 1179

\bibitem[\protect\citeauthoryear{Athanassoula}{Athanassoula}{2012}]{Athanassoula2012}
Athanassoula E.,  2012, \mn@doi [\mnras] {10.1111/j.1745-3933.2012.01320.x},
  426, L46

\bibitem[\protect\citeauthoryear{Athanassoula}{Athanassoula}{2013}]{Athanassoula2013}
Athanassoula E.,  2013, Bars and secular evolution in disk galaxies:
  Theoretical input.
p.~305, \url {http://adsabs.harvard.edu/abs/2013seg..book..305A}

\bibitem[\protect\citeauthoryear{Athanassoula, Bosma  \&
  Papaioannou}{Athanassoula et~al.}{1987}]{Athanassoula1987}
Athanassoula E.,  Bosma A.,   Papaioannou S.,  1987, \aap, 179, 23

\bibitem[\protect\citeauthoryear{Athanassoula, Romero-G{\'o}mez  \&
  Masdemont}{Athanassoula et~al.}{2009}]{Athanassoula2009}
Athanassoula E.,  Romero-G{\'o}mez M.,   Masdemont J.~J.,  2009, \mn@doi
  [\mnras] {10.1111/j.1365-2966.2008.14273.x}, 394, 67

\bibitem[\protect\citeauthoryear{Athanassoula, Romero-G{\'o}mez, Bosma  \&
  Masdemont}{Athanassoula et~al.}{2010}]{Athanassoula2010}
Athanassoula E.,  Romero-G{\'o}mez M.,  Bosma A.,   Masdemont J.~J.,  2010,
  \mn@doi [\mnras] {10.1111/j.1365-2966.2010.17010.x}, 407, 1433

\bibitem[\protect\citeauthoryear{Baba}{Baba}{2015}]{Baba2015}
Baba J.,  2015, \mn@doi [\mnras] {10.1093/mnras/stv2220}, 454, 2954

\bibitem[\protect\citeauthoryear{Baba, Asaki, Makino, Miyoshi, Saitoh  \&
  Wada}{Baba et~al.}{2009}]{Baba2009}
Baba J.,  Asaki Y.,  Makino J.,  Miyoshi M.,  Saitoh T.~R.,   Wada K.,  2009,
  \mn@doi [\apj] {10.1088/0004-637X/706/1/471}, 706, 471

\bibitem[\protect\citeauthoryear{Baba, Saitoh  \& Wada}{Baba
  et~al.}{2013}]{Baba2013}
Baba J.,  Saitoh T.~R.,   Wada K.,  2013, \mn@doi [\apj]
  {10.1088/0004-637X/763/1/46}, 763, 46

\bibitem[\protect\citeauthoryear{Beckman, Font, Borlaff  \&
  Garc{\'{\i}}a-Lorenzo}{Beckman et~al.}{2018}]{Beckman2018}
Beckman J.~E.,  Font J.,  Borlaff A.,   Garc{\'{\i}}a-Lorenzo B.,  2018,
  \mn@doi [\apj] {10.3847/1538-4357/aaa965}, 854, 182

\bibitem[\protect\citeauthoryear{Berentzen, Shlosman, Martinez-Valpuesta  \&
  Heller}{Berentzen et~al.}{2007}]{Berentzen2007}
Berentzen I.,  Shlosman I.,  Martinez-Valpuesta I.,   Heller C.~H.,  2007,
  \mn@doi [\apj] {10.1086/520531}, 666, 189

\bibitem[\protect\citeauthoryear{Berman, Pollard  \& Hockney}{Berman
  et~al.}{1979}]{Berman1979}
Berman R.~H.,  Pollard D.~J.,   Hockney R.~W.,  1979, \aap, 78, 133

\bibitem[\protect\citeauthoryear{Binney \& Tremaine}{Binney \&
  Tremaine}{2008}]{Binney2008}
Binney J.,  Tremaine S.,  2008, Galactic Dynamics: Second Edition.
Princeton University Press

\bibitem[\protect\citeauthoryear{Block, Buta, Knapen, Elmegreen, Elmegreen  \&
  Puerari}{Block et~al.}{2004}]{Block2004}
Block D.~L.,  Buta R.,  Knapen J.~H.,  Elmegreen D.~M.,  Elmegreen B.~G.,
  Puerari I.,  2004, \mn@doi [\aj] {10.1086/421362}, 128, 183

\bibitem[\protect\citeauthoryear{Buta, Knapen, Elmegreen, Salo, Laurikainen,
  Elmegreen, Puerari  \& Block}{Buta et~al.}{2009}]{Buta2009}
Buta R.~J.,  Knapen J.~H.,  Elmegreen B.~G.,  Salo H.,  Laurikainen E.,
  Elmegreen D.~M.,  Puerari I.,   Block D.~L.,  2009, \mn@doi [\aj]
  {10.1088/0004-6256/137/5/4487}, 137, 4487

\bibitem[\protect\citeauthoryear{Buta et~al.,}{Buta et~al.}{2015}]{Buta2015}
Buta R.~J.,  et~al., 2015, \mn@doi [\apjs] {10.1088/0067-0049/217/2/32}, 217,
  32

\bibitem[\protect\citeauthoryear{Cedr{\'e}s, Cepa, Bongiovanni, Casta{\~n}eda,
  S{\'a}nchez-Portal  \& Tomita}{Cedr{\'e}s et~al.}{2013}]{Cedres2013}
Cedr{\'e}s B.,  Cepa J.,  Bongiovanni {\'A}.,  Casta{\~n}eda H.,
  S{\'a}nchez-Portal M.,   Tomita A.,  2013, \mn@doi [\aap]
  {10.1051/0004-6361/201321588}, 560, A59

\bibitem[\protect\citeauthoryear{Collier, Shlosman  \& Heller}{Collier
  et~al.}{2018}]{Collier2018}
Collier A.,  Shlosman I.,   Heller C.,  2018, \mn@doi [\mnras]
  {10.1093/mnras/sty270}, 476, 1331

\bibitem[\protect\citeauthoryear{Considere \& Athanassoula}{Considere \&
  Athanassoula}{1982}]{Considere1982}
Considere S.,  Athanassoula E.,  1982, \aap, 111, 28

\bibitem[\protect\citeauthoryear{Considere \& Athanassoula}{Considere \&
  Athanassoula}{1988}]{Considere1988}
Considere S.,  Athanassoula E.,  1988, \aaps, 76, 365

\bibitem[\protect\citeauthoryear{Contopoulos}{Contopoulos}{1980}]{Contopoulos1980}
Contopoulos G.,  1980, \aap, 81, 198

\bibitem[\protect\citeauthoryear{D'Onghia}{D'Onghia}{2015}]{Donghia2015}
D'Onghia E.,  2015, \mn@doi [\apjl] {10.1088/2041-8205/808/1/L8}, 808, L8

\bibitem[\protect\citeauthoryear{Davis \& Hayes}{Davis \&
  Hayes}{2014}]{Davis2014}
Davis D.~R.,  Hayes W.~B.,  2014, \mn@doi [\apj] {10.1088/0004-637X/790/2/87},
  790, 87

\bibitem[\protect\citeauthoryear{Davis, Berrier, Shields, Kennefick, Kennefick,
  Seigar, Lacy  \& Puerari}{Davis et~al.}{2012}]{Davis2012}
Davis B.~L.,  Berrier J.~C.,  Shields D.~W.,  Kennefick J.,  Kennefick D.,
  Seigar M.~S.,  Lacy C. H.~S.,   Puerari I.,  2012, \mn@doi [\apjs]
  {10.1088/0067-0049/199/2/33}, 199, 33

\bibitem[\protect\citeauthoryear{Davis, Graham  \& Seigar}{Davis
  et~al.}{2017}]{Davis2017}
Davis B.~L.,  Graham A.~W.,   Seigar M.~S.,  2017, \mn@doi [\mnras]
  {10.1093/mnras/stx1794}, 471, 2187

\bibitem[\protect\citeauthoryear{D{\'{\i}}az-Garc{\'{\i}}a \&
  Kanpen}{D{\'{\i}}az-Garc{\'{\i}}a \& Kanpen}{2018}]{Diaz2018}
D{\'{\i}}az-Garc{\'{\i}}a S.,  Kanpen J.~H.,  2018, in preparation

\bibitem[\protect\citeauthoryear{Dobbs \& Baba}{Dobbs \&
  Baba}{2014}]{Dobbs2014}
Dobbs C.,  Baba J.,  2014, \mn@doi [\pasa] {10.1017/pasa.2014.31}, 31, e035

\bibitem[\protect\citeauthoryear{Dobbs \& Pringle}{Dobbs \&
  Pringle}{2010}]{Dobbs2010}
Dobbs C.~L.,  Pringle J.~E.,  2010, \mn@doi [\mnras]
  {10.1111/j.1365-2966.2010.17323.x}, 409, 396

\bibitem[\protect\citeauthoryear{Egusa, Kohno, Sofue, Nakanishi  \&
  Komugi}{Egusa et~al.}{2009}]{Egusa2009}
Egusa F.,  Kohno K.,  Sofue Y.,  Nakanishi H.,   Komugi S.,  2009, \mn@doi
  [\apj] {10.1088/0004-637X/697/2/1870}, 697, 1870

\bibitem[\protect\citeauthoryear{Elmegreen \& Elmegreen}{Elmegreen \&
  Elmegreen}{1985}]{Elmegreen1985}
Elmegreen B.~G.,  Elmegreen D.~M.,  1985, \mn@doi [\apj] {10.1086/162810}, 288,
  438

\bibitem[\protect\citeauthoryear{Epinat, Amram  \& Marcelin}{Epinat
  et~al.}{2008}]{Epinat2008}
Epinat B.,  Amram P.,   Marcelin M.,  2008, \mn@doi [\mnras]
  {10.1111/j.1365-2966.2008.13796.x}, 390, 466

\bibitem[\protect\citeauthoryear{Erwin \& Sparke}{Erwin \&
  Sparke}{2003}]{Erwin2003}
Erwin P.,  Sparke L.~S.,  2003, \mn@doi [\apjs] {10.1086/367885}, 146, 299

\bibitem[\protect\citeauthoryear{Font, Beckman, Epinat, Fathi, Guti{\'e}rrez
  \& Hernandez}{Font et~al.}{2011}]{Font2011}
Font J.,  Beckman J.~E.,  Epinat B.,  Fathi K.,  Guti{\'e}rrez L.,   Hernandez
  O.,  2011, \mn@doi [\apjl] {10.1088/2041-8205/741/1/L14}, 741, L14

\bibitem[\protect\citeauthoryear{Font, Beckman, Querejeta, Epinat, James,
  Blasco-herrera, Erroz-Ferrer  \& P{\'e}rez}{Font et~al.}{2014a}]{Font2014a}
Font J.,  Beckman J.~E.,  Querejeta M.,  Epinat B.,  James P.~A.,
  Blasco-herrera J.,  Erroz-Ferrer S.,   P{\'e}rez I.,  2014a, \mn@doi [\apjs]
  {10.1088/0067-0049/210/1/2}, 210, 2

\bibitem[\protect\citeauthoryear{Font, Beckman, Zaragoza-Cardiel, Fathi, Epinat
   \& Amram}{Font et~al.}{2014b}]{Font2014b}
Font J.,  Beckman J.~E.,  Zaragoza-Cardiel J.,  Fathi K.,  Epinat B.,   Amram
  P.,  2014b, \mn@doi [\mnras] {10.1093/mnrasl/slu120}, 444, L85

\bibitem[\protect\citeauthoryear{Font et~al.,}{Font et~al.}{2017}]{Font2017}
Font J.,  et~al., 2017, \mn@doi [\apj] {10.3847/1538-4357/835/2/279}, 835, 279

\bibitem[\protect\citeauthoryear{Foyle, Rix, Dobbs, Leroy  \& Walter}{Foyle
  et~al.}{2011}]{Foyle2011}
Foyle K.,  Rix H.-W.,  Dobbs C.~L.,  Leroy A.~K.,   Walter F.,  2011, \mn@doi
  [\apj] {10.1088/0004-637X/735/2/101}, 735, 101

\bibitem[\protect\citeauthoryear{Goldreich \& Lynden-Bell}{Goldreich \&
  Lynden-Bell}{1965}]{Goldreich1965}
Goldreich P.,  Lynden-Bell D.,  1965, \mn@doi [\mnras]
  {10.1093/mnras/130.2.125}, 130, 125

\bibitem[\protect\citeauthoryear{Gonzalez \& Graham}{Gonzalez \&
  Graham}{1996}]{Gonzalez1996}
Gonzalez R.~A.,  Graham J.~R.,  1996, \mn@doi [\apj] {10.1086/176999}, 460, 651

\bibitem[\protect\citeauthoryear{Grand, Kawata  \& Cropper}{Grand
  et~al.}{2012}]{Grand2012}
Grand R. J.~J.,  Kawata D.,   Cropper M.,  2012, \mn@doi [\mnras]
  {10.1111/j.1365-2966.2012.21733.x}, 426, 167

\bibitem[\protect\citeauthoryear{Grand, Kawata  \& Cropper}{Grand
  et~al.}{2013}]{Grand2013}
Grand R. J.~J.,  Kawata D.,   Cropper M.,  2013, \mn@doi [\aap]
  {10.1051/0004-6361/201321308}, 553, A77

\bibitem[\protect\citeauthoryear{Grosb{\o}l, Patsis  \& Pompei}{Grosb{\o}l
  et~al.}{2004}]{Grosbol2004}
Grosb{\o}l P.,  Patsis P.~A.,   Pompei E.,  2004, \mn@doi [\aap]
  {10.1051/0004-6361:20035804}, 423, 849

\bibitem[\protect\citeauthoryear{Grosb{\o}l, Dottori  \& Gredel}{Grosb{\o}l
  et~al.}{2006}]{Grosbol2006}
Grosb{\o}l P.,  Dottori H.,   Gredel R.,  2006, \mn@doi [\aap]
  {10.1051/0004-6361:20065446}, 453, L25

\bibitem[\protect\citeauthoryear{Hayes \& Davis}{Hayes \&
  Davis}{2012}]{Hayes2012}
Hayes W.~B.,  Davis D.,  2012, in AAS/Division of Dynamical Astronomy Meeting
  \#43. p.~6.04, \url {http://adsabs.harvard.edu/abs/2012DDA....43.0604H}

\bibitem[\protect\citeauthoryear{Hernandez et~al.,}{Hernandez
  et~al.}{2008}]{Hernandez2008}
Hernandez O.,  et~al., 2008, \mn@doi [\pasp] {10.1086/589844}, 120, 665

\bibitem[\protect\citeauthoryear{Huntley, Sanders  \& Roberts}{Huntley
  et~al.}{1978}]{Huntley1978}
Huntley J.~M.,  Sanders R.~H.,   Roberts Jr. W.~W.,  1978, \mn@doi [\apj]
  {10.1086/156054}, 221, 521

\bibitem[\protect\citeauthoryear{Julian \& Toomre}{Julian \&
  Toomre}{1966}]{Julian1966}
Julian W.~H.,  Toomre A.,  1966, \mn@doi [\apj] {10.1086/148957}, 146, 810

\bibitem[\protect\citeauthoryear{Kalnajs}{Kalnajs}{1975}]{Kalnajs1975}
Kalnajs A.~J.,  1975, in L. Weliachew, Colloq. Internat. CNRS. p.~103

\bibitem[\protect\citeauthoryear{Kalnajs}{Kalnajs}{1978}]{Kalnajs1978}
Kalnajs A.~J.,  1978, in {Berkhuijsen} E.~M.,  {Wielebinski} R.,  eds,  IAU
  Symposium Vol. 77, Structure and Properties of Nearby Galaxies. pp 113--125,
  \url {http://adsabs.harvard.edu/abs/1978IAUS...77..113K}

\bibitem[\protect\citeauthoryear{Kennefick}{Kennefick}{2014}]{Kennefick2014}
Kennefick D.,  2014, in {Seigar} M.~S.,  {Treuthardt} P.,  eds,  Astronomical
  Society of the Pacific Conference Series Vol. 480, Structure and Dynamics of
  Disk Galaxies. p.~125, \url
  {http://adsabs.harvard.edu/abs/2014ASPC..480..125K}

\bibitem[\protect\citeauthoryear{Kennicutt}{Kennicutt}{1981}]{Kennicutt1981}
Kennicutt Jr. R.~C.,  1981, \mn@doi [\aj] {10.1086/113064}, 86, 1847

\bibitem[\protect\citeauthoryear{Kormendy \& Norman}{Kormendy \&
  Norman}{1979}]{Kormendy1979}
Kormendy J.,  Norman C.~A.,  1979, \mn@doi [\apj] {10.1086/157414}, 233, 539

\bibitem[\protect\citeauthoryear{Krajnovi{\'c}, Cappellari, de Zeeuw  \&
  Copin}{Krajnovi{\'c} et~al.}{2006}]{Krajnovic2006}
Krajnovi{\'c} D.,  Cappellari M.,  de Zeeuw P.~T.,   Copin Y.,  2006, \mn@doi
  [\mnras] {10.1111/j.1365-2966.2005.09902.x}, 366, 787

\bibitem[\protect\citeauthoryear{Laurikainen, Salo, Buta  \&
  Knapen}{Laurikainen et~al.}{2007}]{Laurikainen2007}
Laurikainen E.,  Salo H.,  Buta R.,   Knapen J.~H.,  2007, \mn@doi [\mnras]
  {10.1111/j.1365-2966.2007.12299.x}, 381, 401

\bibitem[\protect\citeauthoryear{Lin \& Lau}{Lin \& Lau}{1975}]{Lin1975}
Lin C.~C.,  Lau Y.~Y.,  1975, SIAM Journal of Applied Mathematics, 29, 352

\bibitem[\protect\citeauthoryear{Lin \& Shu}{Lin \& Shu}{1964}]{Lin1964}
Lin C.~C.,  Shu F.~H.,  1964, \mn@doi [\apj] {10.1086/147955}, 140, 646

\bibitem[\protect\citeauthoryear{Lynden-Bell}{Lynden-Bell}{1979}]{Linden1979}
Lynden-Bell D.,  1979, \mn@doi [\mnras] {10.1093/mnras/187.1.101}, 187, 101

\bibitem[\protect\citeauthoryear{Martin}{Martin}{1995}]{Martin1995}
Martin P.,  1995, \mn@doi [\aj] {10.1086/117461}, 109, 2428

\bibitem[\protect\citeauthoryear{Mart{\'{\i}}nez-Garc{\'{\i}}a}{Mart{\'{\i}}nez-Garc{\'{\i}}a}{2012}]{Martinez2012}
Mart{\'{\i}}nez-Garc{\'{\i}}a 2012, \apj, 744, 92

\bibitem[\protect\citeauthoryear{Mart{\'{\i}}nez-Garc{\'{\i}}a,
  Gonz{\'a}lez-L{\'o}pezlira  \& Bruzual-A}{Mart{\'{\i}}nez-Garc{\'{\i}}a
  et~al.}{2009}]{Martinez2009}
Mart{\'{\i}}nez-Garc{\'{\i}}a E.~E.,  Gonz{\'a}lez-L{\'o}pezlira R.~A.,
  Bruzual-A G.,  2009, \mn@doi [\apj] {10.1088/0004-637X/694/1/512}, 694, 512

\bibitem[\protect\citeauthoryear{Martinez-Valpuesta, Shlosman  \&
  Heller}{Martinez-Valpuesta et~al.}{2006}]{Martinez2006}
Martinez-Valpuesta I.,  Shlosman I.,   Heller C.,  2006, \mn@doi [\apj]
  {10.1086/498338}, 637, 214

\bibitem[\protect\citeauthoryear{Masset \& Tagger}{Masset \&
  Tagger}{1997}]{Masset1997}
Masset F.,  Tagger M.,  1997, \aap, 322, 442

\bibitem[\protect\citeauthoryear{Mata-Ch{\'a}vez, G{\'o}mez  \&
  Puerari}{Mata-Ch{\'a}vez et~al.}{2014}]{Mata-Chavez2014}
Mata-Ch{\'a}vez M.~D.,  G{\'o}mez G.~C.,   Puerari I.,  2014, \mn@doi [\mnras]
  {10.1093/mnras/stu1672}, 444, 3756

\bibitem[\protect\citeauthoryear{Minchev, Famaey, Quillen, Di~Matteo, Combes,
  Vlaji{\'c}, Erwin  \& Bland-Hawthorn}{Minchev et~al.}{2012}]{Minchev2012}
Minchev I.,  Famaey B.,  Quillen A.~C.,  Di~Matteo P.,  Combes F.,  Vlaji{\'c}
  M.,  Erwin P.,   Bland-Hawthorn J.,  2012, \mn@doi [\aap]
  {10.1051/0004-6361/201219198}, 548, A126

\bibitem[\protect\citeauthoryear{Ohta, Hamabe  \& Wakamatsu}{Ohta
  et~al.}{1990}]{Ohta1990}
Ohta K.,  Hamabe M.,   Wakamatsu K.-I.,  1990, \mn@doi [\apj] {10.1086/168892},
  357, 71

\bibitem[\protect\citeauthoryear{{\^O}ki, Fujimoto  \& Hitotuyanagi}{{\^O}ki
  et~al.}{1964}]{Oki1964}
{\^O}ki T.,  Fujimoto M.,   Hitotuyanagi Z.,  1964, \mn@doi [Progress of
  Theoretical Physics Supplement] {10.1143/PTPS.31.77}, 31, 77

\bibitem[\protect\citeauthoryear{Patsis}{Patsis}{2006}]{Patsis2006}
Patsis P.~A.,  2006, \mn@doi [\mnras] {10.1111/j.1745-3933.2006.00174.x}, 369,
  L56

\bibitem[\protect\citeauthoryear{Patsis, Hiotelis, Contopoulos  \&
  Grosbol}{Patsis et~al.}{1994}]{Patsis1994}
Patsis P.~A.,  Hiotelis N.,  Contopoulos G.,   Grosbol P.,  1994, \aap, 286, 46

\bibitem[\protect\citeauthoryear{Patsis, Grosbol  \& Hiotelis}{Patsis
  et~al.}{1997}]{Patsis1997}
Patsis P.~A.,  Grosbol P.,   Hiotelis N.,  1997, \aap, 323, 762

\bibitem[\protect\citeauthoryear{Puerari \& Dottori}{Puerari \&
  Dottori}{1992}]{Puerari1992}
Puerari I.,  Dottori H.~A.,  1992, \aaps, 93, 469

\bibitem[\protect\citeauthoryear{Puerari, Block, Elmegreen, Frogel  \&
  Eskridge}{Puerari et~al.}{2000}]{Puerari2000}
Puerari I.,  Block D.~L.,  Elmegreen B.~G.,  Frogel J.~A.,   Eskridge P.~B.,
  2000, \aap, 359, 932

\bibitem[\protect\citeauthoryear{Puerari, Elmegreen  \& Block}{Puerari
  et~al.}{2014}]{Puerari2014}
Puerari I.,  Elmegreen B.~G.,   Block D.~L.,  2014, \mn@doi [\aj]
  {10.1088/0004-6256/148/6/133}, 148, 133

\bibitem[\protect\citeauthoryear{Rautiainen \& Salo}{Rautiainen \&
  Salo}{1999}]{Rautiainen1999}
Rautiainen P.,  Salo H.,  1999, \aap, 348, 737

\bibitem[\protect\citeauthoryear{Ringermacher \& Mead}{Ringermacher \&
  Mead}{2009}]{Ringermacher2009}
Ringermacher H.~I.,  Mead L.~R.,  2009, \mn@doi [\mnras]
  {10.1111/j.1365-2966.2009.14950.x}, 397, 164

\bibitem[\protect\citeauthoryear{Roberts, Roberts  \& Shu}{Roberts
  et~al.}{1975}]{Roberts1975}
Roberts Jr. W.~W.,  Roberts M.~S.,   Shu F.~H.,  1975, \mn@doi [\apj]
  {10.1086/153421}, 196, 381

\bibitem[\protect\citeauthoryear{Roberts, Huntley  \& van Albada}{Roberts
  et~al.}{1979}]{Roberts1979}
Roberts Jr. W.~W.,  Huntley J.~M.,   van Albada G.~D.,  1979, \mn@doi [\apj]
  {10.1086/157367}, 233, 67

\bibitem[\protect\citeauthoryear{Romero-G{\'o}mez, Masdemont, Athanassoula  \&
  Garc{\'{\i}}a-G{\'o}mez}{Romero-G{\'o}mez et~al.}{2006}]{Romero-Gomez2006}
Romero-G{\'o}mez M.,  Masdemont J.~J.,  Athanassoula E.,
  Garc{\'{\i}}a-G{\'o}mez C.,  2006, \mn@doi [\aap]
  {10.1051/0004-6361:20054653}, 453, 39

\bibitem[\protect\citeauthoryear{Romero-G{\'o}mez, Athanassoula, Masdemont  \&
  Garc{\'{\i}}a-G{\'o}mez}{Romero-G{\'o}mez et~al.}{2007}]{Romero-Gomez2007}
Romero-G{\'o}mez M.,  Athanassoula E.,  Masdemont J.~J.,
  Garc{\'{\i}}a-G{\'o}mez C.,  2007, \mn@doi [\aap]
  {10.1051/0004-6361:20077504}, 472, 63

\bibitem[\protect\citeauthoryear{Ro{\v s}kar, Debattista, Quinn  \&
  Wadsley}{Ro{\v s}kar et~al.}{2012}]{Roskar2012}
Ro{\v s}kar R.,  Debattista V.~P.,  Quinn T.~R.,   Wadsley J.,  2012, \mn@doi
  [\mnras] {10.1111/j.1365-2966.2012.21860.x}, 426, 2089

\bibitem[\protect\citeauthoryear{Sanders}{Sanders}{1977}]{Sanders1977}
Sanders R.~H.,  1977, \mn@doi [\apj] {10.1086/155637}, 217, 916

\bibitem[\protect\citeauthoryear{Sanders \& Huntley}{Sanders \&
  Huntley}{1976}]{Sanders1976}
Sanders R.~H.,  Huntley J.~M.,  1976, \mn@doi [\apj] {10.1086/154692}, 209, 53

\bibitem[\protect\citeauthoryear{Savchenko \& Reshetnikov}{Savchenko \&
  Reshetnikov}{2013}]{Savchenko2013}
Savchenko S.~S.,  Reshetnikov V.~P.,  2013, \mn@doi [\mnras]
  {10.1093/mnras/stt1627}, 436, 1074

\bibitem[\protect\citeauthoryear{Schempp}{Schempp}{1982}]{Schempp1982}
Schempp W.~V.,  1982, \mn@doi [\apj] {10.1086/160056}, 258, 96

\bibitem[\protect\citeauthoryear{Schlosser \& Musculus}{Schlosser \&
  Musculus}{1984}]{Schlosser1984}
Schlosser W.,  Musculus D.,  1984, \aap, 131, 367

\bibitem[\protect\citeauthoryear{Seigar \& James}{Seigar \&
  James}{1998}]{Seigar1998}
Seigar M.~S.,  James P.~A.,  1998, \mn@doi [\mnras]
  {10.1046/j.1365-8711.1998.01779.x}, 299, 685

\bibitem[\protect\citeauthoryear{Seigar, Bullock, Barth  \& Ho}{Seigar
  et~al.}{2006}]{Seigar2006}
Seigar M.~S.,  Bullock J.~S.,  Barth A.~J.,   Ho L.~C.,  2006, \mn@doi [\apj]
  {10.1086/504463}, 645, 1012

\bibitem[\protect\citeauthoryear{Seigar, Kennefick, Kennefick  \& Lacy}{Seigar
  et~al.}{2008}]{Seigar2008}
Seigar M.~S.,  Kennefick D.,  Kennefick J.,   Lacy C. H.~S.,  2008, \mn@doi
  [\apjl] {10.1086/588727}, 678, L93

\bibitem[\protect\citeauthoryear{Sellwood}{Sellwood}{2014}]{Sellwood2014}
Sellwood J.~A.,  2014, \mn@doi [Reviews of Modern Physics]
  {10.1103/RevModPhys.86.1}, 86, 1

\bibitem[\protect\citeauthoryear{Sellwood \& Sparke}{Sellwood \&
  Sparke}{1988}]{Sellwood1988}
Sellwood J.~A.,  Sparke L.~S.,  1988, \mn@doi [\mnras]
  {10.1093/mnras/231.1.25P}, 231, 25P

\bibitem[\protect\citeauthoryear{Sellwood \& Spekkens}{Sellwood \&
  Spekkens}{2015}]{Sellwood2015}
Sellwood J.~A.,  Spekkens K.,  2015, preprint (\mn@eprint {arXiv} {1509.07120})

\bibitem[\protect\citeauthoryear{Sheth et~al.,}{Sheth et~al.}{2010}]{Sheth2010}
Sheth K.,  et~al., 2010, \mn@doi [\pasp] {10.1086/657638}, 122, 1397

\bibitem[\protect\citeauthoryear{Shields et~al.,}{Shields
  et~al.}{2015}]{Shields2015}
Shields D.~W.,  et~al., 2015, preprint (\mn@eprint {arXiv} {1511.06365})

\bibitem[\protect\citeauthoryear{Shu}{Shu}{2016}]{Shu2016}
Shu F.~H.,  2016, \mn@doi [\araa] {10.1146/annurev-astro-081915-023426}, 54,
  667

\bibitem[\protect\citeauthoryear{Sormani, Binney  \& Magorrian}{Sormani
  et~al.}{2015}]{Sormani2015}
Sormani M.~C.,  Binney J.,   Magorrian J.,  2015, \mn@doi [\mnras]
  {10.1093/mnras/stv1135}, 451, 3437

\bibitem[\protect\citeauthoryear{Sygnet, Tagger, Athanassoula  \&
  Pellat}{Sygnet et~al.}{1988}]{Sygnet1988}
Sygnet J.~F.,  Tagger M.,  Athanassoula E.,   Pellat R.,  1988, \mn@doi
  [\mnras] {10.1093/mnras/232.4.733}, 232, 733

\bibitem[\protect\citeauthoryear{Tagger, Sygnet, Athanassoula  \&
  Pellat}{Tagger et~al.}{1987}]{Tagger1987}
Tagger M.,  Sygnet J.~F.,  Athanassoula E.,   Pellat R.,  1987, \mn@doi [\apjl]
  {10.1086/184934}, 318, L43

\bibitem[\protect\citeauthoryear{Tamburro, Rix, Walter, Brinks, de Blok,
  Kennicutt  \& Mac~Low}{Tamburro et~al.}{2008}]{Tamburro2008}
Tamburro D.,  Rix H.-W.,  Walter F.,  Brinks E.,  de Blok W. J.~G.,  Kennicutt
  R.~C.,   Mac~Low M.-M.,  2008, \mn@doi [\aj] {10.1088/0004-6256/136/6/2872},
  136, 2872

\bibitem[\protect\citeauthoryear{Toomre \& Kalnajs}{Toomre \&
  Kalnajs}{1991}]{Toomre1991}
Toomre A.,  Kalnajs A.~J.,  1991, in {Sundelius} B.,  ed., Dynamics of Disc
  Galaxies. p.~341, \url {http://adsabs.harvard.edu/abs/1991dodg.conf..341T}

\bibitem[\protect\citeauthoryear{Tremaine \& Weinberg}{Tremaine \&
  Weinberg}{1984}]{Tremaine1984}
Tremaine S.,  Weinberg M.~D.,  1984, \mn@doi [\apjl] {10.1086/184292}, 282, L5

\bibitem[\protect\citeauthoryear{Tsoutsis, Efthymiopoulos  \& Voglis}{Tsoutsis
  et~al.}{2008}]{Tsoutsis2008}
Tsoutsis P.,  Efthymiopoulos C.,   Voglis N.,  2008, \mn@doi [\mnras]
  {10.1111/j.1365-2966.2008.13331.x}, 387, 1264

\bibitem[\protect\citeauthoryear{Tsoutsis, Kalapotharakos, Efthymiopoulos  \&
  Contopoulos}{Tsoutsis et~al.}{2009}]{Tsoutsis2009}
Tsoutsis P.,  Kalapotharakos C.,  Efthymiopoulos C.,   Contopoulos G.,  2009,
  \mn@doi [\aap] {10.1051/0004-6361:200810149}, 495, 743

\bibitem[\protect\citeauthoryear{Villa-Vargas, Shlosman  \&
  Heller}{Villa-Vargas et~al.}{2009}]{Villa2009}
Villa-Vargas J.,  Shlosman I.,   Heller C.,  2009, \mn@doi [\apj]
  {10.1088/0004-637X/707/1/218}, 707, 218

\bibitem[\protect\citeauthoryear{Villa-Vargas, Shlosman  \&
  Heller}{Villa-Vargas et~al.}{2010}]{Villa2010}
Villa-Vargas J.,  Shlosman I.,   Heller C.,  2010, \mn@doi [\apj]
  {10.1088/0004-637X/719/2/1470}, 719, 1470

\bibitem[\protect\citeauthoryear{Voglis, Tsoutsis  \& Efthymiopoulos}{Voglis
  et~al.}{2006}]{Voglis2006}
Voglis N.,  Tsoutsis P.,   Efthymiopoulos C.,  2006, \mn@doi [\mnras]
  {10.1111/j.1365-2966.2006.11021.x}, 373, 280

\bibitem[\protect\citeauthoryear{Wada}{Wada}{1994}]{Wada1994}
Wada K.,  1994, \pasj, 46, 165

\bibitem[\protect\citeauthoryear{Wozniak \& Pierce}{Wozniak \&
  Pierce}{1991}]{Wozniak1991}
Wozniak H.,  Pierce M.~J.,  1991, \aaps, 88, 325

\bibitem[\protect\citeauthoryear{de Vaucouleurs, de Vaucouleurs, Corwin, Buta,
  Paturel  \& Fouque}{de~Vaucouleurs et~al.}{1995}]{Vaucouleurs1995}
de Vaucouleurs G.,  de Vaucouleurs A.,  Corwin H.~G.,  Buta R.~J.,  Paturel G.,
    Fouque P.,  1995, VizieR Online Data Catalog, 7155

\bibitem[\protect\citeauthoryear{del R{\'{\i}}o \& Cepa}{del R{\'{\i}}o \&
  Cepa}{1999}]{Rio1999}
del R{\'{\i}}o M.~S.,  Cepa J.,  1999, \mn@doi [\aaps] {10.1051/aas:1999440},
  134, 333

\makeatother
\end{thebibliography}



\bsp	
\label{lastpage}
\end{document}